\documentclass[useAMS,usenatbib]{mnras}

\usepackage{newtxtext,newtxmath}

\usepackage[T1]{fontenc}
\usepackage{ae,aecompl}

\usepackage{graphicx}	
\usepackage{amsmath}	
\usepackage{amssymb}	
\usepackage{epstopdf}
\usepackage{subcaption}
\usepackage{soul}

\usepackage[usenames, dvipsnames]{color}

\usepackage{etoolbox}

\newcommand{\teff}{$T_{\rm{eff}}$}
\newcommand{\threeD}{$\langle$3D$\rangle$}
\newcommand{\logg}{$\log{g}$}
\newcommand{\y}{$\log{\rm{(H/He)}}$}
\newcommand{\co}{CO$^5$BOLD}
\newcommand{\taur}{$\log{\tau_{\rm{R}}}$}
\newcommand{\mlt}{ML2/$\alpha$}
\newcommand{\mltf}{ML2/$\alpha_{\rm{f}}$}
\newcommand{\mlts}{ML2/$\alpha_{\rm{S}}$}
\newcommand{\mltjump}{ML2/$\alpha_{\rm{s_{\rm{jump}}}}$}
\newcommand{\mltfmax}{ML2/$\alpha_{\rm{F_{\rm{max}}}}$}
\newcommand{\q}{$\log{(M_{\rm{CVZ}}/M_{\rm{tot}})}$}
\newcommand{\s}{Schwarzschild}
\newcommand{\msun}{$M_{\odot}$}
\newcommand{\vrms}{$v_{z, \rm{rms}}$}
\newcommand{\wdz}{SDSS J073842.56+183509.06}

\title[Calibration of MLT for DB and DBA white dwarfs]{Calibration of the mixing length theory for structures of helium-dominated atmosphere white dwarfs}

\author[Cukanovaite et al.]{E. Cukanovaite$^1$\,{\Huge \footnotemark},
P.-E. Tremblay$^1$, B. Freytag$^2$, H.-G. Ludwig$^3$, G. Fontaine$^4$, 
  \newauthor{P. Brassard$^4$, O. Toloza$^1$ and D. Koester$^5$}
  \\
$^{1}$ Department of Physics, University of Warwick, Coventry CV4 7AL, UK \\ 
$^{2}$ Department of Physics and Astronomy, Uppsala University, Box 516, 751 20 Uppsala, Sweden \\
$^{3}$ Zentrum f\"ur Astronomie der Universit\"at Heidelberg, Landessternwarte, K\"onigstuhl 12, 69117 Heidelberg, Germany \\
$^{4}$ D\'epartement de Physique, Universit\'e de Montr\'eal, C.P. 6128, Succ. Centre-Ville, Montr\'eal, QC H3C 3J7, Canada \\
$^{5}$ Institut f\"ur Theoretische Physik und Astrophysik, Universit\"at Kiel, 24098 Kiel, Germany
}

\date{Accepted XXX. Received YYY; in original form ZZZ}

\pubyear{2019}

\begin{document}
\label{firstpage}
\pagerange{\pageref{firstpage}--\pageref{lastpage}}
\maketitle

\begin{abstract}
We perform a calibration of the mixing length parameter at the bottom boundary of the convection zone for helium-dominated atmospheres of white dwarfs. This calibration is based on a grid of 3D DB (pure-helium) and DBA (helium-dominated with traces of hydrogen) model atmospheres computed with the \co~radiation-hydrodynamics code, and a grid of 1D DB and DBA envelope structures. The 3D models span a parameter space of hydrogen-to-helium abundances between $-10.0\leq$~\y~$\leq -2.0$, surface gravities between 7.5~$\leq$~\logg~$\leq$~9.0 and effective temperatures between 12\,000 K~$\lesssim$~\teff~$\lesssim$~34\,000 K. The 1D envelopes cover a similar atmospheric parameter range, but are also calculated with different values of the mixing length parameter, namely $0.4\leq$~\mlt~$\leq 1.4$. The calibration is performed based on two definitions of the bottom boundary of the convection zone, the \s~and the zero convective flux boundaries. Thus, our calibration is relevant for applications involving the bulk properties of the convection zone including its total mass, which excludes the spectroscopic technique. Overall, the calibrated \mlt~is smaller than what is commonly used in evolutionary models and theoretical determinations of the blue edge of the instability strip for pulsating DB and DBA stars. With calibrated \mlt~we are able to deduce more accurate convection zone sizes needed for studies of planetary debris mixing and dredge-up of carbon from the core. We highlight this by calculating examples of metal-rich 3D DBAZ models and finding their convection zone masses. Mixing length calibration represents the first step of in-depth investigations of convective overshoot in white dwarfs with helium-dominated atmospheres. 
\end{abstract}


\footnotetext{E-mail: E.Cukanovaite@warwick.ac.uk}



\section{Introduction}

Any main-sequence star below $\approx$ 8\msun~will end its life by expelling the majority of its outer envelope and leaving behind a dense, degenerate core, known as a white dwarf \citep{althaus2010}. Due to their large surface gravities (abbreviated as the logarithm of surface gravity, \logg), compositionally these stellar remnants are well-stratified, with the heavier material sinking into the core and the outer layers being composed of the lightest chemical elements present \citep{schatzman48}. In magnitude-limited samples around 80\% of all white dwarfs have hydrogen-dominated atmospheres and 20\% have helium-dominated atmospheres \citep{kleinman13,kepler15}. White dwarfs are unable to fuse matter in their degenerate cores and thus evolve simply by cooling. As they cool, superficial convection zones develop in their envelopes and grow bigger with decreasing effective temperature, \teff~\citep{tassoul1990}. This means that both the structure and evolutionary models of white dwarfs can be affected by uncertainties arising from the treatment of convective energy transport. 

Until recently, the standard white dwarf models used for the atmosphere and the interior have been 1D, where convection is treated using the ML2 version \citep{tassoul1990} of the mixing length theory, MLT \citep{bohm1958}. The formulation of this theory assumes same-sized, large convective eddies travelling a distance, $d$, which is known as the mixing length, before dissipating into the surroundings by releasing (or absorbing) their excess (or deficient) energy. The distance travelled depends on a free parameter called the mixing length parameter, $\alpha$ (or \mlt~to indicate the use of ML2 version of MLT for white dwarfs), such that
\begin{equation}
d = \alpha H_{\rm{p}}~,
\end{equation}
where $H_{\rm{p}}$ is the pressure scale height. This free parameter is not given by the MLT and instead must be calibrated from observations, which is a significant shortcoming of the theory as the particular value of the parameter can have a significant effect on the modelled structures (see examples for both evolutionary and atmospheric models: \citealt{shipman1979,winget1982,winget1983,fontaine1984,tassoul1990,Thejll1991,bergeron1992,koester1994,bergeron1995,wesemael1999,corsico2016}), especially when convection becomes superadiabatic \citep{tremblay2015,sonoi_mlt_evolution}. 

As an improvement, another 1D theory of convection, CMT \citep{canuto1991,canuto1992} and its refined version CGM \citep{canuto1996}, have also been used in modelling white dwarf evolution \citep{althaus1996,althaus1997,Benvenuto1999}. Unlike MLT, CMT does not rely on the approximation of single-sized convective eddies and instead considers a full range of eddy sizes. Unfortunately, similarly to MLT, CMT depends on the local conditions of the atmosphere \citep{ludwig1999}, which is a restrictive approximation as convection is a non-local process. This assumption was subsequently removed in non-local 1D envelope models of white dwarfs \citep{montgomery2004}. Given that convection is inherently a 3D process, the dimensionality issue was first improved by 2D atmospheric models of DA white dwarfs developed by \cite{ludwig1993}, \cite{ludwig_1994_op_binning} and \cite{freytag1996}.

More recently, the first 3D models for pure-hydrogen atmosphere (DA) \citep{tremblay_2013_3dmodels,tremblay_2013_granulation,tremblay_2013_spectra,kupka2018} and pure-helium atmosphere (DB) \citep{myprecious} white dwarfs have been developed. In 3D models convection is non-local, is treated from first principles and the models do not depend on any free parameters, although numerical parameters do exist. Spectroscopic corrections derived from 3D models have been tested against \textit{Gaia} DR2 data \citep{gaia2018} by comparing the observed parallaxes for samples of DA and DB/DBA white dwarfs with spectroscopically-derived parallaxes with and without 3D corrections \citep{tremblay2019}. 3D DA corrections were shown to be in excellent agreement with the data. For the DB/DBA samples, the 3D DB corrections were not a clear improvement upon predicted 1D parallaxes. Given that the 3D corrections were for DB white dwarfs only and the samples contained a large fraction of DBA stars, it was concluded that 3D DBA spectroscopic corrections, as well as a re-evaluation of the line broadening parameters \citep{gb2019}, are needed to proceed. This will be the subject of a future study. 

In this paper, we instead focus on \mlt~calibration at the bottom of the convection zone for 3D DB and DBA models, similar to what has been achieved for 3D DA models \citep{tremblay2015}. We use a new grid of 3D DBA models consisting of 235 simulations alongside the recently published grid of 47 3D DB models. Our calibration of \mlt~is relevant for the overall thermal and mixing properties of the convection zone. It differs in purpose to the \mlt~calibration based on a detailed spectroscopic analysis performed by \cite{bergeron_db_2011}. This is because the spectral light forming layers for DB and DBA stars are always near or above the top of the convection zone. Additionally, due to the dynamic nature of convection, the mixing length parameter varies throughout the white dwarf structure \citep{ludwig_1994_op_binning,tremblay2015}. Therefore, no single 1D synthetic spectrum at a given \mlt~value can reproduce the entirety of a 3D spectrum \citep{myprecious}.

Our calibration is of relevance to many applications. First of all, it is not currently possible to compute 3D evolutionary models of any star. Instead, 1D stellar evolution models have been improved by calibrating the mixing length parameter based on 3D atmospheric models and allowing it to vary accordingly as the star evolves \citep{trampedach_mlt_evolution, magic2015, salaris_mlt_evolution, mosumgaard_mlt_evolution, sonoi_mlt_evolution}. Such calibration has already been performed for DA white dwarfs \citep{tremblay2015}, but has not been done for DB and DBA stars.

The position of the theoretical blue edge of the instability strip for V777 Her (DBV) white dwarfs is heavily dependent on the assumed convective efficiency at the bottom of the convection zone \citep{fontaine2008,corsico2009,vangrootel2017}. Larger \mlt~values result in larger \teff~of the blue edge. The current empirical blue edge of the strip is defined by PG0112+104 at \teff~$\approx$~31\,000~K (at \logg~$\approx$~7.8) \citep{shipman2002,provencal2003,hermes2017}, approximately 2\,000~K higher than the current theoretical blue edge of \teff~$\approx$~29\,000~K (at \logg~$\approx$~7.8) calculated at the spectroscopically-calibrated \mlt~$=1.25$ \citep{vangrootel2017}. This suggests that higher convective efficiency is needed to correctly model the empirical blue edge. 

\mlt~calibration at the bottom of the convection zone can also provide more accurate convection zone sizes for DB and DBA white dwarfs. This is needed in order to understand the accretion of planetesimals onto white dwarfs, including the mixing of the different accreted chemical elements within the convection zone and their diffusion at its bottom (or floating in the case of hydrogen). These events are frequent around DB and DBA white dwarfs \citep{kleinman13, veras16} and could explain the origin of hydrogen in DBA stars \citep{gentilefusillo2017}. However, for a full 3D description of the accretion-diffusion scenario, convective overshoot must also be accounted for \citep{kupka2018,cunningham2019}, which is outside the scope of the current work.

In Sect.~\ref{sec:num} we present the grids of 3D DB and DBA atmospheric models and 1D envelope structures used for the calibration of the \mlt~parameter. Sect.~\ref{sec:conv} describes the general properties of the 3D convection zones and the differences to 1D convection zones. The calibration method is described in Sect.~\ref{sec:cal} and results are discussed in Sect.~\ref{sec:disc}. We conclude in Sect.~\ref{sec:sum}. 

\newpage

\section{Numerical setup}~\label{sec:num}

\subsection{3D atmospheric models}

\begin{figure*}
	\includegraphics[width=1.5\columnwidth]{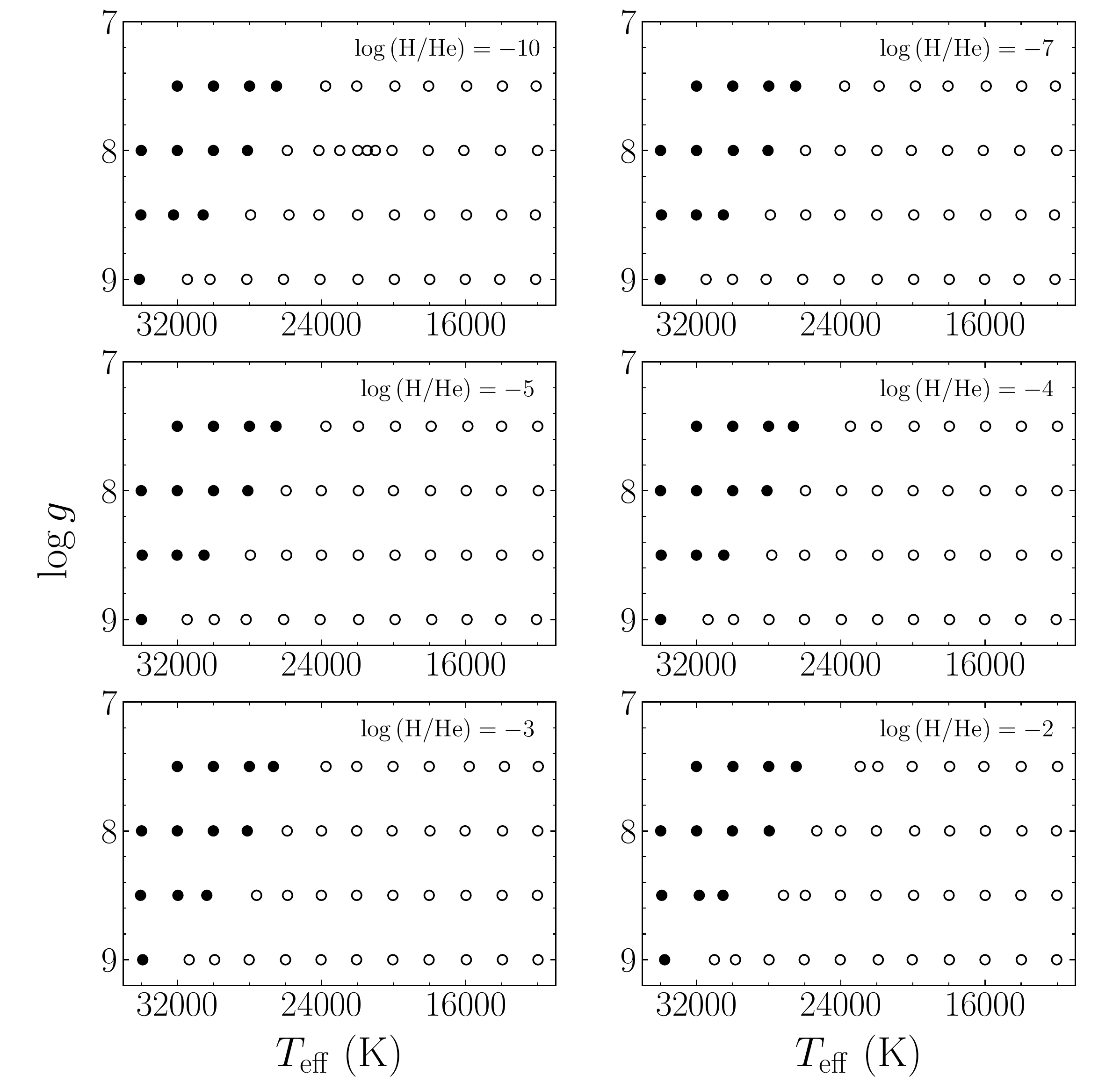}
    \caption{The abundances, surface gravities and effective temperatures of the 3D models presented in this paper. Open and filled circles denote the models with open and closed bottom boundaries, respectively.}
    \label{fig:models}
\end{figure*}

Using the \co~radiation-hydrodynamics code \citep{freytag2002_co,wedemeyer2004_co,freytag2012_cobold,freytag2013,freytag2017}, we have calculated 285 3D DB and DBA models with 12\,000~K~$\lesssim$~\teff~$\lesssim$~34\,000~K, 7.5~$\leq$~\logg~$\leq$~9.0 and $-10.0\leq$~\y~$\leq-2.0$, where \y~is the logarithm of the ratio of the number of hydrogen-to-helium atoms in the atmosphere. Fig.~\ref{fig:models} illustrates the atmospheric parameter values of our 3D simulations. Appendix~\ref{ap:tables} in the Supplementary Material also lists basic information about the 3D models, including their atmospheric parameters, simulation box sizes, running times and intensity contrasts. For DB models we use \y~$=-10.0$ as this low hydrogen abundance practically describes a pure-helium composition. The abundance range chosen covers the majority of observed hydrogen abundances in DB/DBA samples \citep{bergeron_db_2011,koester_kepler_2015,rolland2018}. For all abundances, \logg~=~7.5 models only extend up to 32\,000~K due to convective energy transport being negligible at higher \teff~for this particular \logg. Currently, there are no known low-mass helium-dominated atmosphere white dwarfs, which would be formed as a consequence of binary evolution~\citep{tremblay2019,gb2019}. Therefore, we do not calculate models with \logg~$<7.5$.

The 3D DB simulations have already been presented in \cite{myprecious}. The same numerical setup was used to calculate 3D DBA models but with equations of state (EOS) and opacity tables appropriate for the given hydrogen abundance. More detail on the numerical setup can therefore be found in \cite{myprecious}. In summary, each model is computed using the box-in-a-star \co~setup \citep{freytag2012_cobold}, where a portion of the atmosphere is modelled in a Cartesian 3D box of typical size 150 $\times$ 150 $\times$ 150 ($x \times y \times z$) grid points with $z$ being the geometric height pointing towards the exterior of the white dwarf. Each simulation has periodic side boundaries. The top boundary is always open to material and radiative flows, whereas the bottom boundary can be open or closed to convective flows. For most of our models the convection zone sizes are vertically too large to be simulated. In this case the open bottom boundary is used. As the effective temperature increases, the convection zone shrinks until its vertical size becomes small enough to fit within the simulation box. For these models we use closed bottom boundary where the vertical velocity is forced to go to zero at the boundary. For all simulations the top boundary is located at \taur~$\lesssim -5.0$, where \taur~is the logarithm of the Rosseland optical depth. The bottom boundaries are around \taur~$=3.0$, however, some closed bottom simulations had to be extended deeper to justify the enforcement of zero vertical velocity. In most extreme cases, the models had to be vertically extended to 230 grids points, increasing \taur~to around 4. 

For a given model the input parameters are an equation of state, an opacity table, \logg~and a parameter that controls the \teff~of the model. The \teff~value is recovered after the simulation is run from the spatially and temporally averaged emergent flux. In the case of open bottom models, the entropy of the inflowing material at the bottom boundary controls the \teff. For closed bottom models, the controlling parameter is the radiative flux specified at the bottom. For all abundances we use opacity tables with 10 bins with boundaries at \taur~$=$~[99.0, 0.25, 0.0, $-$0.25, $-$0.5, $-$1.0, $-$1.5, $-$2.0, $-$3.0, $-$4.0, $-$5.0] based on reference 1D models. We rely on the binning technique as outlined in \cite{nordlund_1982_opac_binning}, \cite{ludwig_1994_op_binning}, \cite{vogler_2004_op_binning} and \cite{myprecious}. We do not include the far-UV opacities assigned to the~[$-$5.0, $-$99.0] bin due to interpolation issues as was the case for 3D DB simulations \citep{myprecious}. The opacity tables and EOS are based on the 1D models of \cite{bergeron_db_2011}, which include the Stark profiles of neutral helium from \cite{beauchamp_stark_profiles_1997} and the free-free absorption coefficient of negative helium ions from \cite{john_neg_ab_coeff_neg_he_ion_1994}. For DBA models the Stark broadening of \cite{tremblay_stark_2009} is used for hydrogen lines.

The 3D models are spatially- and temporally-averaged in order to extract the relevant atmospheric stratifications, i.e. entropy, temperature, pressure and convective flux as functions of \taur. The spatial average is performed over constant geometric height, unlike in \cite{myprecious} where the spatial average was done over contours of constant \taur. The temporal average is performed over the last quarter of the simulation, i.e. the last quarter of the total run time given in Tabs.~\ref{tab:3d_models10}-\ref{tab:3d_models2}. We confirm that our models are relaxed by monitoring the total flux at all depths and the convergence of the velocity field \citep{myprecious}. Relaxation usually occurs in the first half of the simulation, as we start from a simulation that is already close to the final solution.

\subsection{1D envelope models}~\label{sec:oneDmodels}

In order to find a mixing length value that best matches the nature of 3D convection zones, we use the updated 1D DB and DBA envelope models of \cite{vangrootel2017} and \cite{fontaine2001}, which span the same parameter range as our 3D atmospheric models but also different values of \mlt, namely 0.4~$\leq$~\mlt~$\leq$~1.4 in steps of 0.1. The envelopes rely on non-grey upper boundary conditions extracted from the atmospheric models of \cite{bergeron_db_2011}, and on the non-ideal EOS of \cite{saumon1995}. Turbulent pressure is not included in the envelope structures.

\begin{figure*}
	\includegraphics[width=1.5\columnwidth]{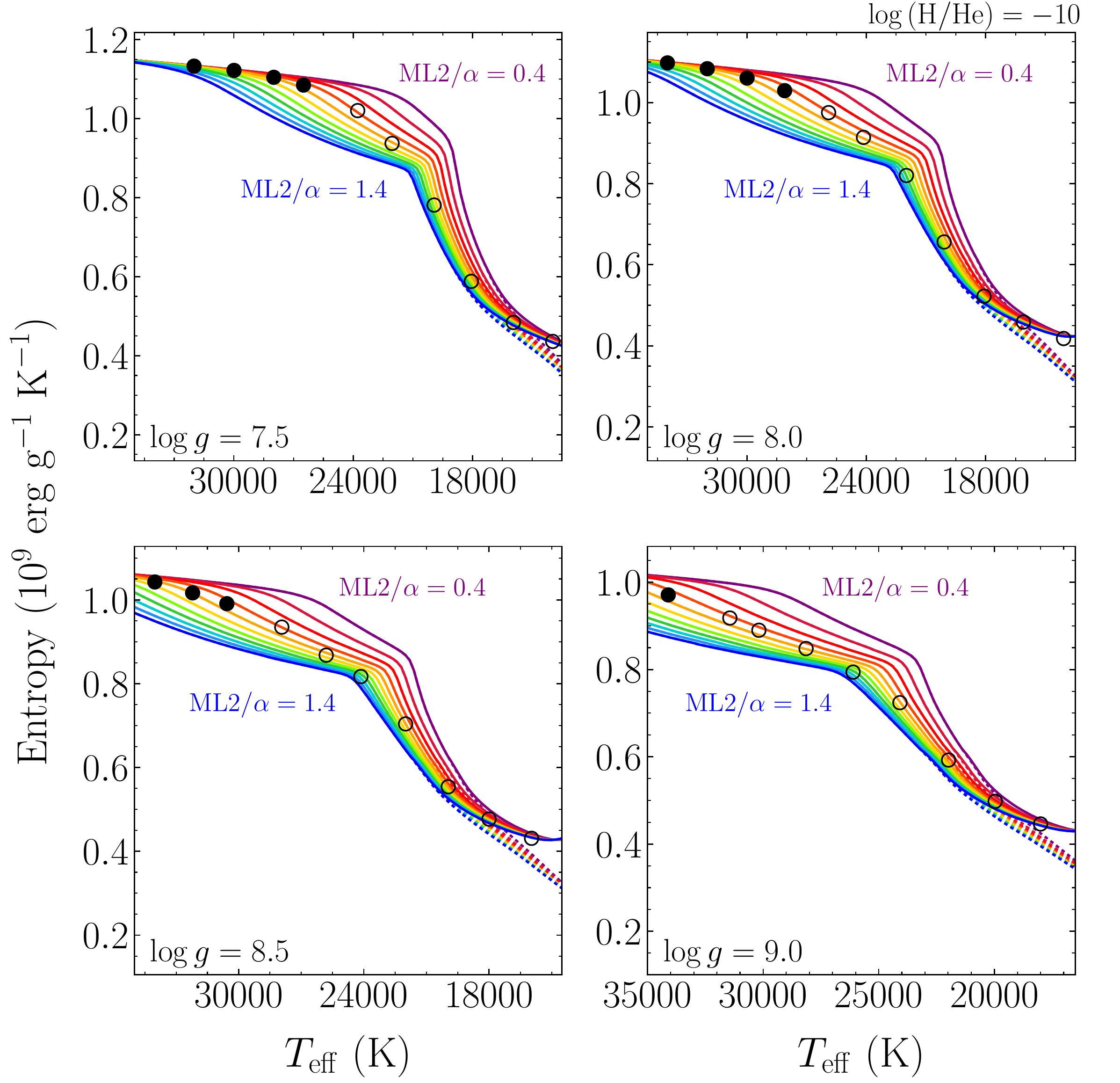}
    \caption{The entropy at the bottom of the convection zone defined by the \s~criterion as a function of \teff~for 3D DB open (open circles) and closed (filled circles) bottom models, and for 1D DB envelopes with different values of the mixing length parameter. The \mlt~value decreases by increments of 0.1 from the dark blue line (\mlt~$=1.4$) all the way up to the dark purple line (\mlt~$=~0.4$). We show the 1D entropies with (solid lines) and without (dashed lines) partial degeneracy effects taken into account. The \logg~values of the models are indicated on the panels.}
    \label{fig:s_teff}
\end{figure*}

For the majority of 3D models the inflowing entropy at the base of the convection zone (the input parameter for open bottom models which controls \teff~of the model) is used for \mlt~calibration. In order to have a common entropy zero-point between the 1D envelopes and 3D atmospheres, we re-calculate the 1D entropy from temperature and pressure at the base of the 1D envelope convection zone. The entropy is re-calculated with and without partial degeneracy to demonstrate the degeneracy effects. Fig.~\ref{fig:s_teff} shows entropy as a function of \teff~for selected models. At high \teff~the partial degeneracy is negligible as the chemical potential of free electrons has a large negative value. Partial degeneracy becomes important for cool \teff~models due to their low temperatures and high densities. For the \y~$=-10.0$ grid, our first-order partial degeneracy correction begins to break down for the lowest \teff~models not plotted in Fig.~\ref{fig:s_teff}, namely \teff~$ \lesssim 14\,000$, $14\,000$, $16\,000$, $18\,000$ K for \logg~$=7.5$, 8.0, 8.5, 9.0 models, respectively. Similar behaviour is observed for the DBA grid. Below these \teff~convection in envelopes is almost fully adiabatic everywhere and becomes independent of the particular choice of \mlt. Therefore, we do not attempt calibration of \mlt~in that particular \teff~regime (see Sect.~\ref{sec:cal}). We find that partial degeneracy is more important for low \teff~DB/DBA models than low \teff~DA models (see Fig. 1 of \citealt{tremblay2015}) possibly due to the higher densities of DB models. 

\begin{figure*}
	\includegraphics[width=1.5\columnwidth]{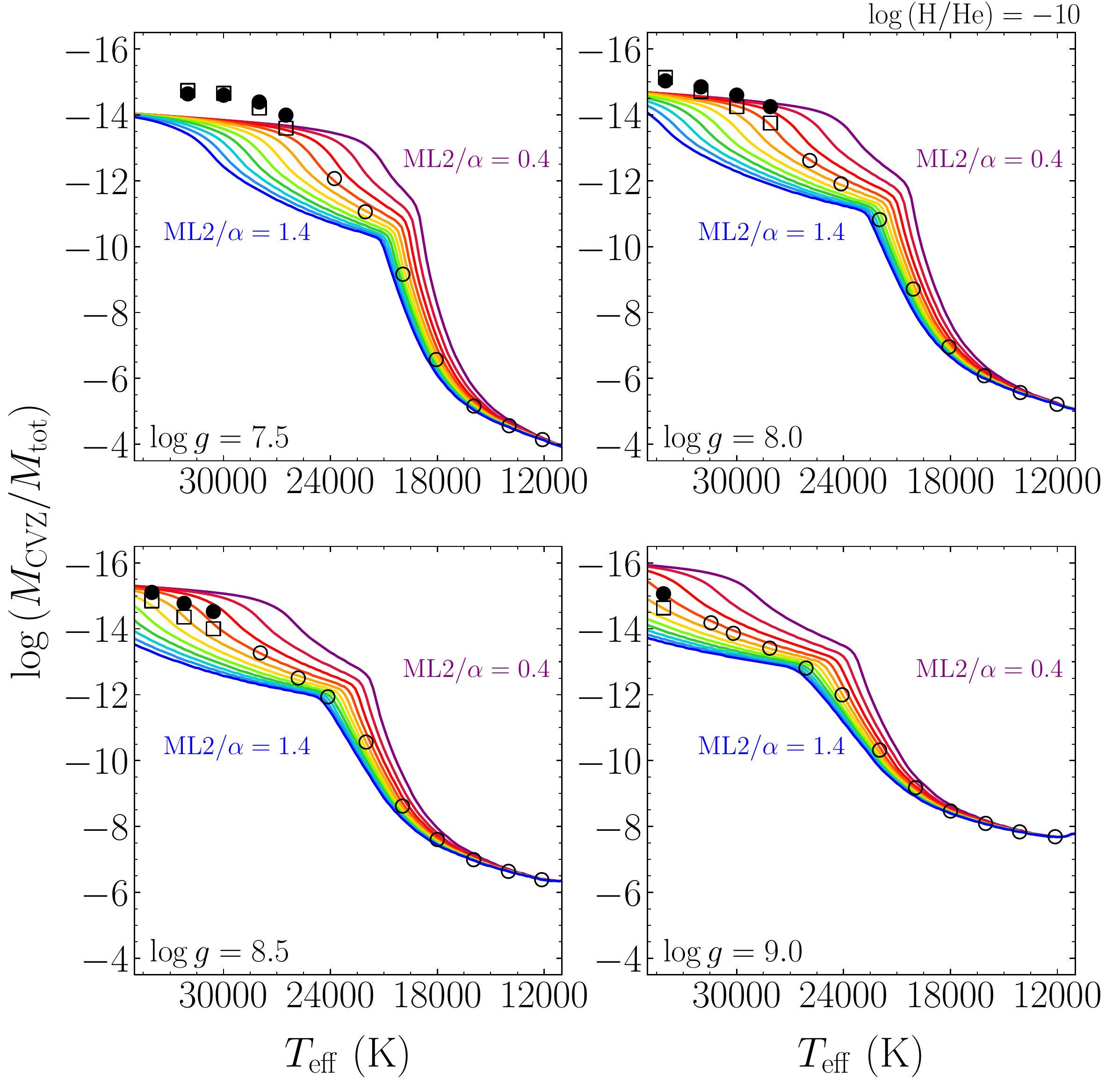}
    \caption{The fraction of the convection zone mass to the total mass of the white dwarf as a function of \teff~for 3D DB models and 1D DB envelopes (solid lines) with different values of the mixing length parameter. The \mlt~value decreases by increments of 0.1 from the dark blue line (\mlt~$=1.4$) all the way up to the dark purple line (\mlt~$=0.4$). The \s~boundaries for the 3D open bottom models are indicated by open circles; filled circles represent the Schwarzschild boundary for closed bottom 3D models; open squares represent the flux boundary for closed bottom 3D models.}
    \label{fig:q_teff}
\end{figure*}

From 1D envelopes we also extract the ratio \q, where $M_{\rm{CVZ}}$ is the mass of the convection zone integrated from the surface of the white dwarf to the bottom of the convection zone and $M_{\rm{tot}}$ is the total mass of the white dwarf. An example of this is shown in Fig.~\ref{fig:q_teff}. As expected, varying the value of the \mlt~parameter for models where superadiabatic convection is important has a significant effect on the mass of the convection zone. The change can be as much as $\approx 4$ dex for \logg~=~7.5 DB and DBA models and $\approx 3$ dex for \logg~=~9.0 models. By calibrating \mlt~with our 3D models (see Sect.~\ref{sec:cal}) we can narrow down the uncertainty on the mixed mass within the convection zone.

\begin{figure*}
	\includegraphics[width=1.5\columnwidth]{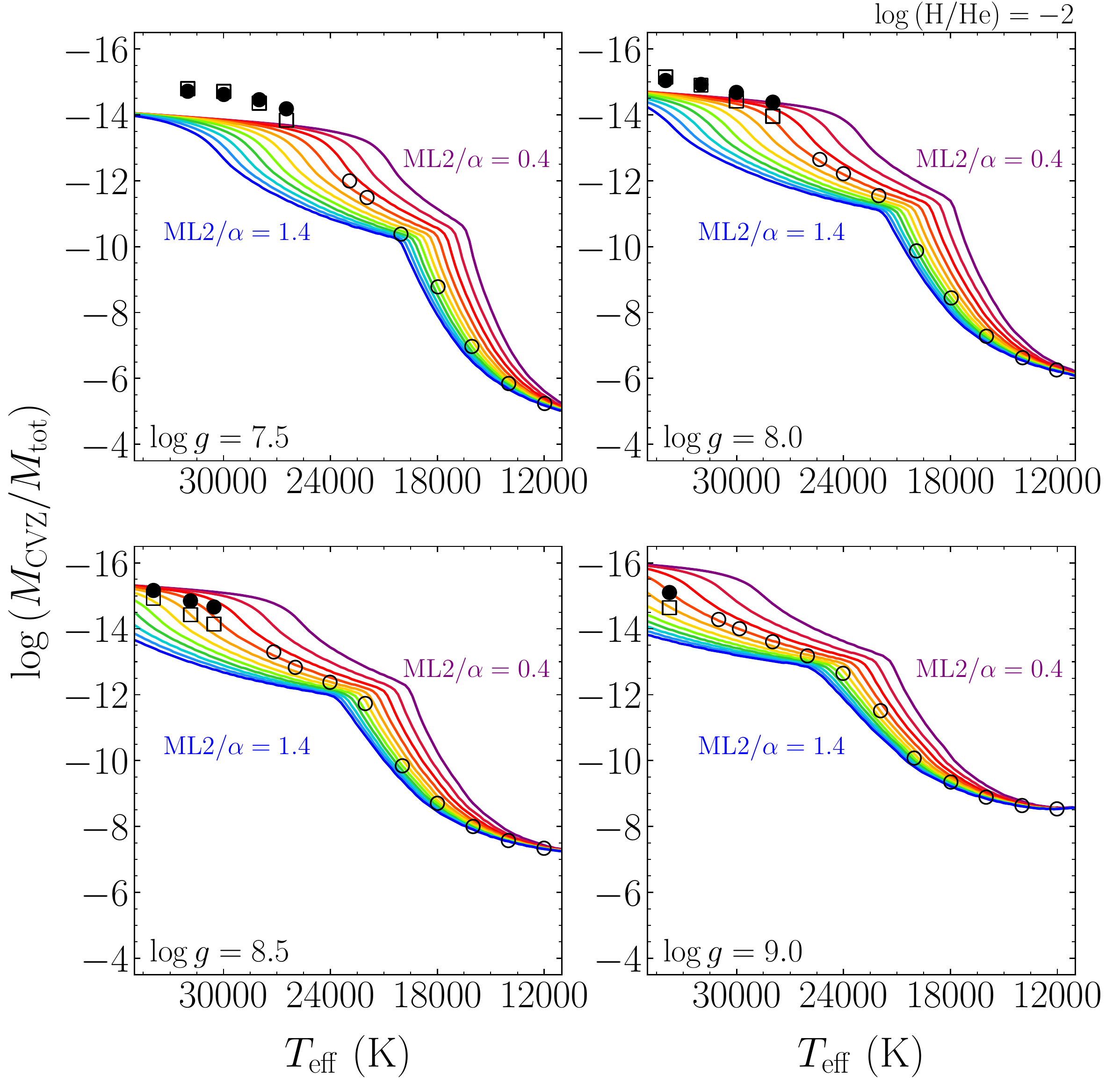}
    \caption{Same as Fig.~\ref{fig:q_teff}, but for a DBA grid with \y~$= -2.0$.}
    \label{fig:q_teff_y-2}
\end{figure*}

The convection zone size increases with decreasing \logg~and decreasing \teff~\citep{fontaine1976}. Shallower convection zones are expected for DBA models as the presence of hydrogen increases the total opacity, decreasing the atmospheric density and pressure \citep{fontaine1976}. This is also seen for late-type stars with increased metallicity \citep{magic2013}. The decrease in density and pressure results in higher adiabatic entropy (see Sec.~\ref{sec:conv}), and therefore lower convective efficiency (and entropy jump, see Sec.~\ref{sec:sjump}) and smaller convection zones \citep{magic2013}. Fig.~\ref{fig:q_teff_y-2} shows \q~for the \y~$=-2.0$ grid. By comparing Figs.~\ref{fig:q_teff} and~\ref{fig:q_teff_y-2} it is clear that the presence of hydrogen does indeed shrink the convection zones.

\section{The convection zone}~\label{sec:conv}

The envelopes of cool DA and DB white dwarfs are convective, with the top of the convection zone almost perfectly overlapping with the photospheric layers \citep{tassoul1990}, meaning that convection is essential for modelling both atmospheres and envelopes of cool white dwarfs. In 1D atmospheric and envelope models the convective layers are defined by the \s~criterion 
\begin{equation}
\left(\frac{\partial \ln{T}}{\partial \ln{P}} \right)_{\rm{radiative}} > \left(\frac{\partial \ln{T}}{\partial \ln{P}} \right)_{\rm{adiabatic}},
\end{equation}
where $T$ and $P$ are the temperature and pressure. Therefore, only those layers that locally satisfy this inequality are able to transport energy through convection, leading to abrupt and clearly-defined boundaries of the convection zone in 1D. This is a limited approximation of the turbulent nature of convection, which is better explored with the use of 3D models. There are at least two ways one can define convection zone boundaries and subsequently convection zone sizes in 3D simulations. In the following we use the \s~criterion (the \s~boundary) and the zero convective flux (the flux boundary) definitions. 

The \s~criterion can be rewritten in terms of the entropy gradient with respect to \taur, such that the convective layers are defined by
\begin{equation}
\frac{\rm{d} \textit{s}}{\rm{d} \tau_{\rm{R}}} > 0~,
\end{equation}
where $s$ is the entropy. We use this definition to determine the edges of the convection zone in both 1D and \threeD~entropy stratifications, focusing on the bottom boundary, defining it to be the \s~boundary.

Unlike in the 1D case, the 3D convective energy is transported even beyond the \s~boundary. This is due to the acceleration of the overdense convective downdrafts in the layers just above the base of the convection zone. In response, because of mass conservation warm material is transported upwards, resulting in a positive convective flux \citep{tremblay2015}.  We define the flux boundary to be the region where the ratio of convective-to-total flux goes to zero. The convective flux, $F_{\rm{conv}}$, is calculated using 
\begin{equation} \label{eq:fconv}
F_{\rm{conv}} = \left \langle \left( e_{\rm{int}} + \frac{P}{\rho} \right) \rho u_z \right\rangle + \left \langle \frac{\mathbf{u}^2}{2} \rho u_z\right \rangle - e_{\rm{tot}} \langle \rho u_z \rangle,
\end{equation}
where $e_{\rm{int}}$ is the internal energy per gram, $\rho$ is the density, $u_z$ is the vertical velocity, $\mathbf{u}$ is the velocity vector and $e_{\rm{tot}}$ is the total energy, defined as 
\begin{equation}
e_{\rm{tot}} = \frac{ \langle \rho e_{\rm{int}} + P + \rho \frac{\mathbf{u}^2 }{2} \rangle }{\langle \rho \rangle}.
\end{equation} 
The first term of Eq.~\ref{eq:fconv} is the enthalpy flux, the second term is the kinetic energy flux and the third term is the mass flux weighted energy flux, which is subtracted in order to correct for any non-zero mass flux arising in the numerical simulations. This definition is identical to the one used in \cite{tremblay2015}. Some authors, for instance \cite{cattaneo1991} and \cite{canuto2007}, have referred to the sum of enthalpy and kinetic energy flux as "convected" flux. In general, convective flux is a synonym for enthalpy flux only. By adding kinetic energy flux, the "convective flux" boundary is moved closer to the \s~boundary, as kinetic energy is always negative for simulations presented here, which have standard granulation topology of slow and broad upflows surrounded by fast and narrow downflows. Therefore, \mlt~values calibrated based on the enthalpy and kinetic flux boundary will be smaller than the calibrated values based on enthalpy flux alone \citep{kupka2018,tremblay2015}. As shown by \cite{kupka2018} the boundary associated with the enthalpy flux indicates where downflows become hotter than their surroundings, which is related to buoyancy, the driving mechanism of convection. Therefore, the definition of convective flux based on enthalpy flux would be crucial in studies of downflows. However, for consistency with previous work of \cite{tremblay2015} we use the definition of "convective" flux as defined in Eq.~\ref{eq:fconv}. In MLT, convective flux refers to enthalpy flux only, as kinetic flux is zero everywhere.

\begin{figure}
	\includegraphics[width=\columnwidth]{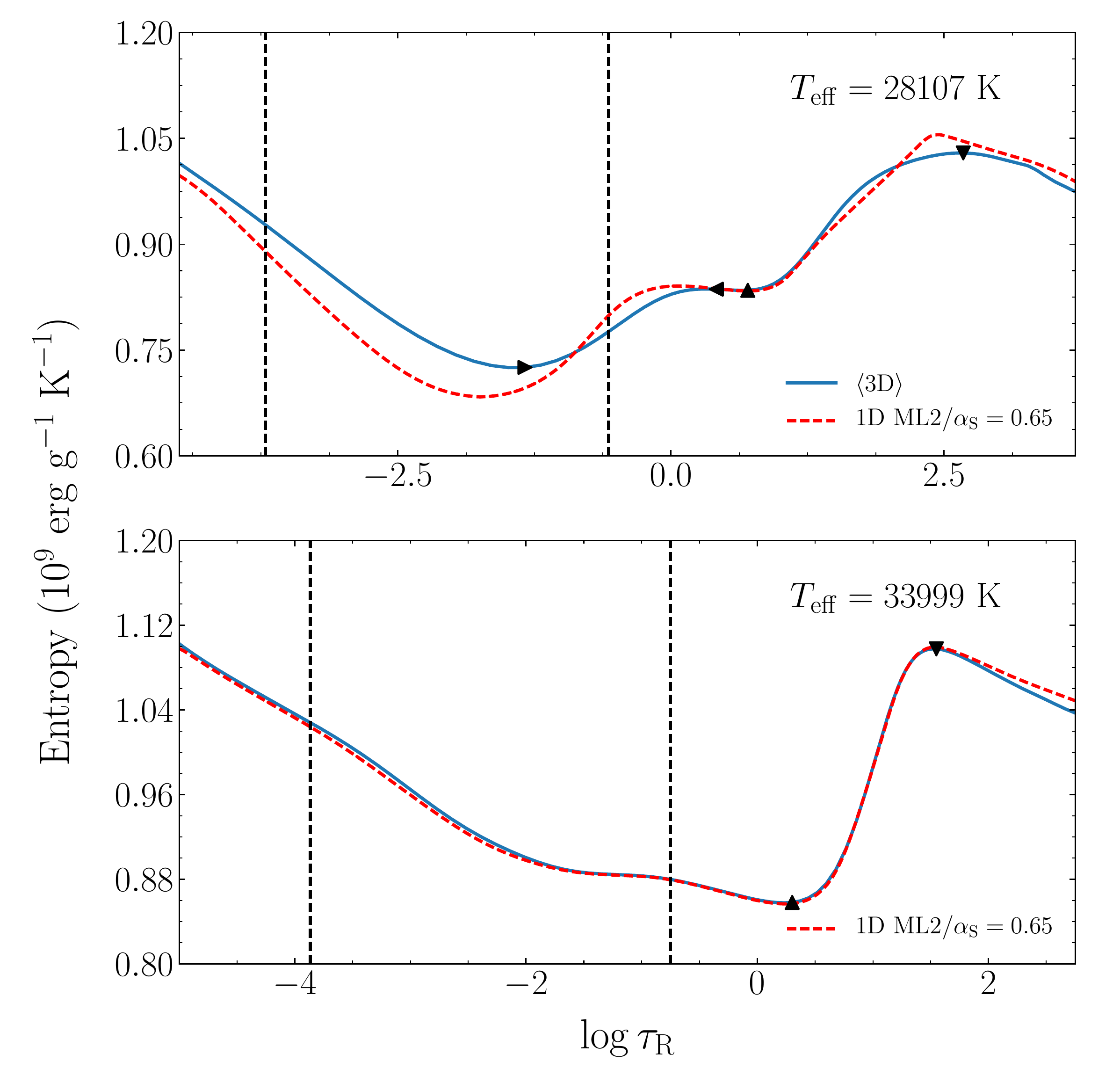}
    \caption{Entropy stratifications of two 3D closed bottom models with \logg~$= 8.0$ and \y~$=-10.0$ are shown as solid blue lines. The dashed black lines indicate the flux-forming region for wavelengths 3500~\AA~to 7200~\AA, representing the atmosphere of the white dwarf in terms of visible light. 1D models calculated at calibrated \mlts~are shown as dashed red lines. According to the \s~criterion, at \teff~$\approx$ 28\,000~K there are two convectively unstable regions due to He~I and He~II ionization. The top and bottom of the first convective region is denoted by right- and left-pointing triangles, respectively. The second convective region is indicated by upward- and downward-pointing triangles. The two convective regions are separated by a small region which is convectively stable in terms of the \s~criterion. At \teff~$\approx$ 34\,000 K, according to the \s~criterion there is only one convective region (He~II) left, which is denoted by the upward- and downward-pointing triangles.}
    \label{fig:closed_bottom_s}
\end{figure}

\begin{figure}
	\includegraphics[width=\columnwidth]{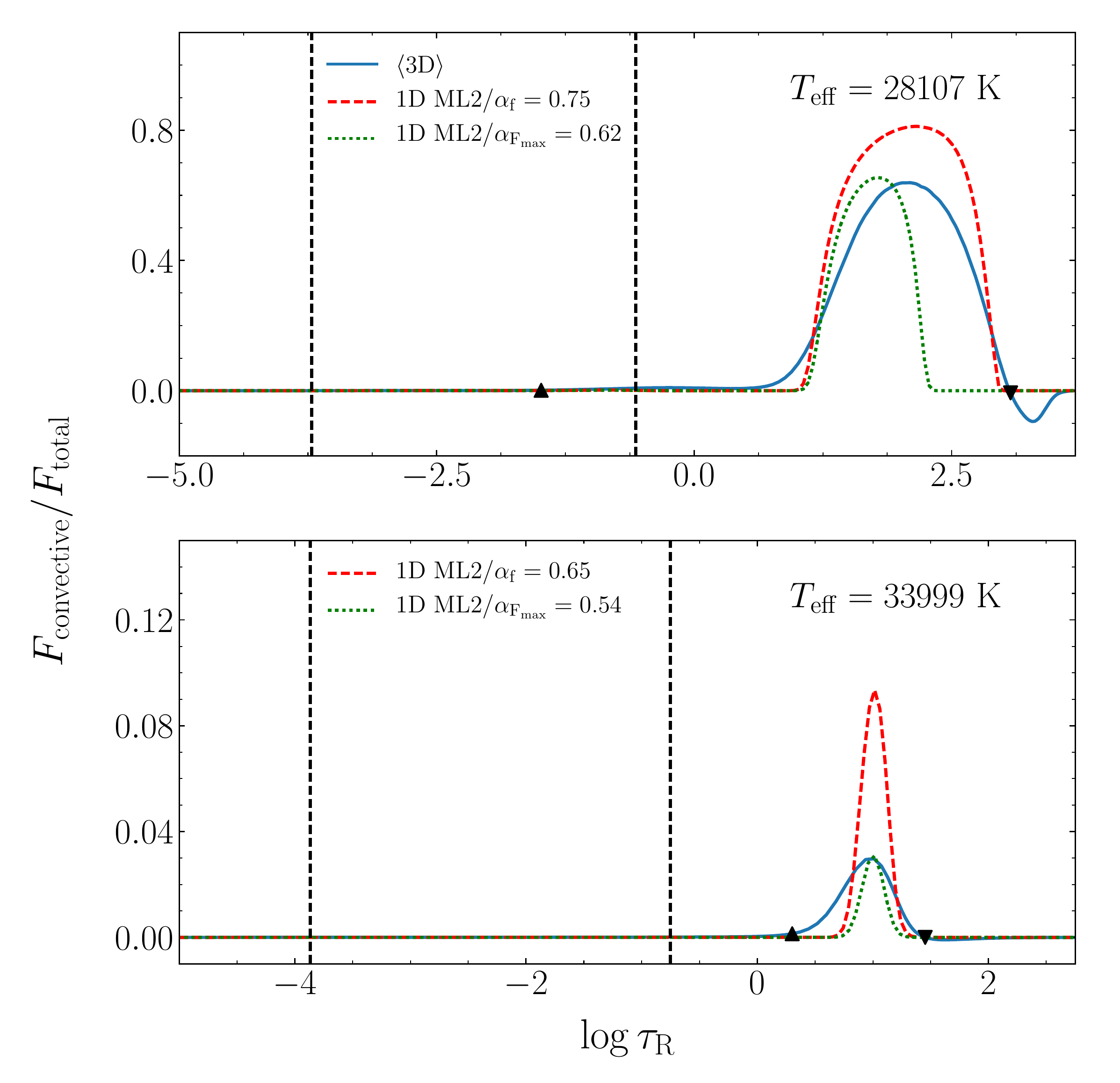}
    \caption{The ratio of the convective-to-total flux as a function of the \taur~for two 3D closed bottom \logg~$=8.0$, \y~$=-10.0$ models is shown in solid blue. The upward- and downward-pointing triangles denote the top and bottom flux boundaries of the convection zone, respectively. The dashed black lines represent the flux-forming region for wavelengths 3500~\AA~to 7200~\AA. Red dashed lines show the 1D models calculated at calibrated \mltf, and green dotted lines show 1D models calculated at \mltfmax~(see Sect.~\ref{sec:fmax}). Unlike the \s~boundary, at \teff~$\approx$ 28\,000~K the two convectively-unstable regions are inseparable in terms of the flux due to the dynamics of the downdrafts. Beyond the flux boundary, a region of negative flux related to convective overshoot is observed.}
    \label{fig:closed_bottom_hcon}
\end{figure}

Figs.~\ref{fig:closed_bottom_s} and~\ref{fig:closed_bottom_hcon} demonstrate the \s~and flux boundaries, respectively. In the case of helium-dominated atmosphere white dwarfs, at higher \teff~there are two convectively-unstable regions related to He~I and He~II ionization. These zones can either be separated by a convectively stable region or merge into one convection zone depending on the \teff. This can also happen for a model at the same \teff, but for different definitions of the convection zone as shown in Figs.~\ref{fig:closed_bottom_s} and~\ref{fig:closed_bottom_hcon}, where the model at \teff~$\approx$~28\,000 K has two clearly defined and separated convectively-unstable regions in terms of the \s~criterion, yet in terms of the flux criterion the two helium zones are indistinguishable, since the flux boundary penetrates deeper. At the highest \teff~only the He~II convection zone remains as He~I is fully ionised. 

In Fig.~\ref{fig:closed_bottom_hcon} we see a region beyond the flux boundary where the ratio of convective-to-total flux becomes negative. This is the convective overshoot region, where the negative convective flux is due to the convective downflow plumes being warmer than the surroundings \citep{zahn1991,tremblay2015}. There is no equivalent region in 1D models and therefore we do not attempt to calibrate the mixing length in any form to describe this region. However, overshoot is important for convective mixing studies. For DA white dwarfs it has been shown that more material can be mixed in the convection zone even beyond the negative flux region (the velocity overshoot region), impacting the mass, abundances, and diffusion times of accreted metals \citep{freytag1996,koester2009,kupka2018,cunningham2019}. This is still unexplored for helium-rich atmospheres.

\section{The calibration method}~\label{sec:cal}

\subsection{Closed bottom models}~\label{sec:closed}

\begin{figure*}
	\includegraphics[width=1.5\columnwidth]{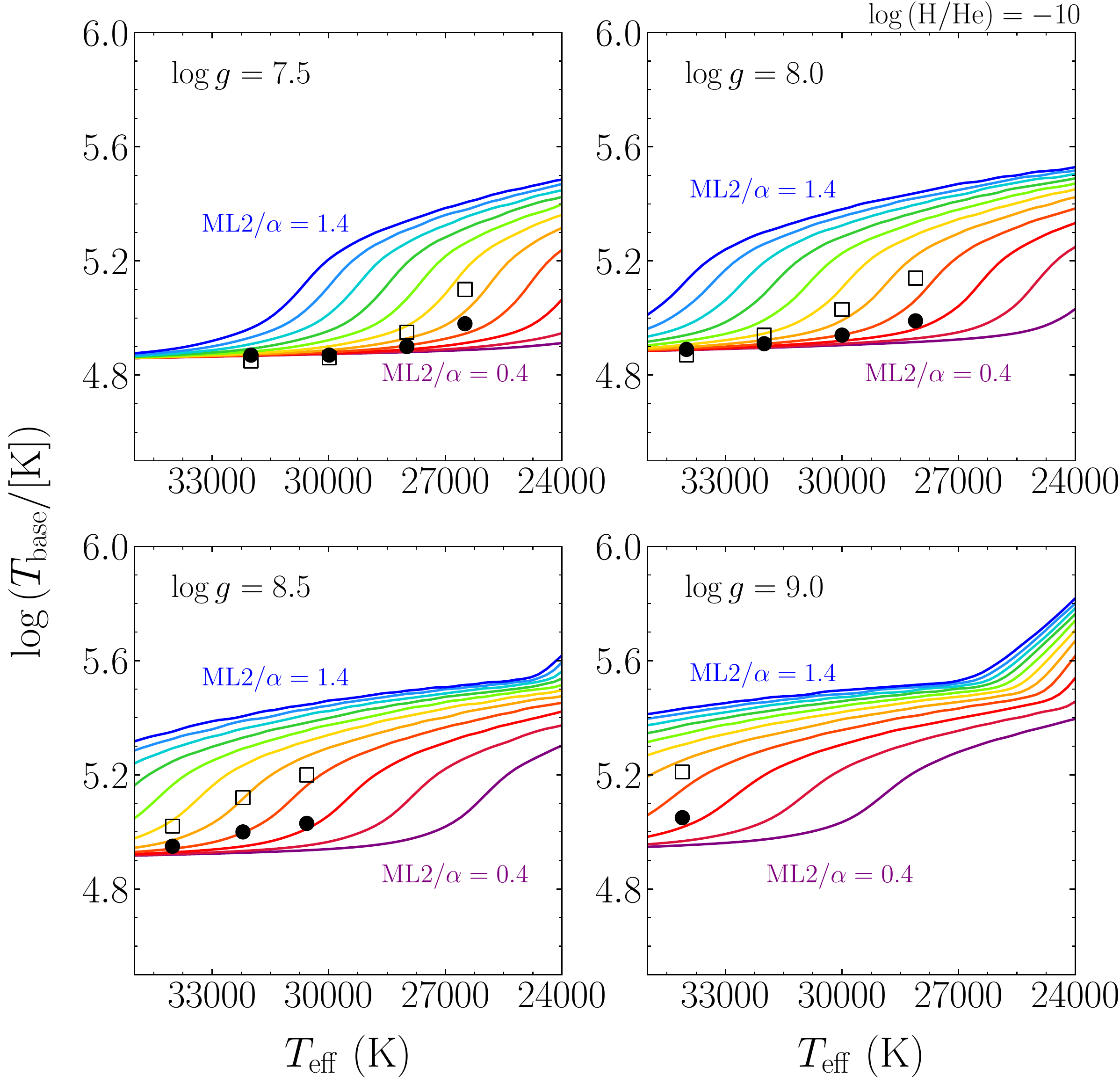}
    \caption{The logarithm of the temperature at the base of the convection zone as a function of \teff~for DB white dwarfs. The solid lines are 1D envelope temperatures at the \s~boundary for varying \mlt~values. The \mlt~value decreases by increments of 0.1 from the dark blue line (\mlt~$=1.4$) all the way down to the dark purple line (\mlt~$=0.4$). The solid circles represent the temperature of closed bottom 3D models at the \s~boundary, the open squares are the temperatures of closed bottom 3D models at the flux boundary. The \logg~values are indicated on the plots.}
    \label{fig:t_closed}
\end{figure*}

\begin{figure*}
	\includegraphics[width=1.5\columnwidth]{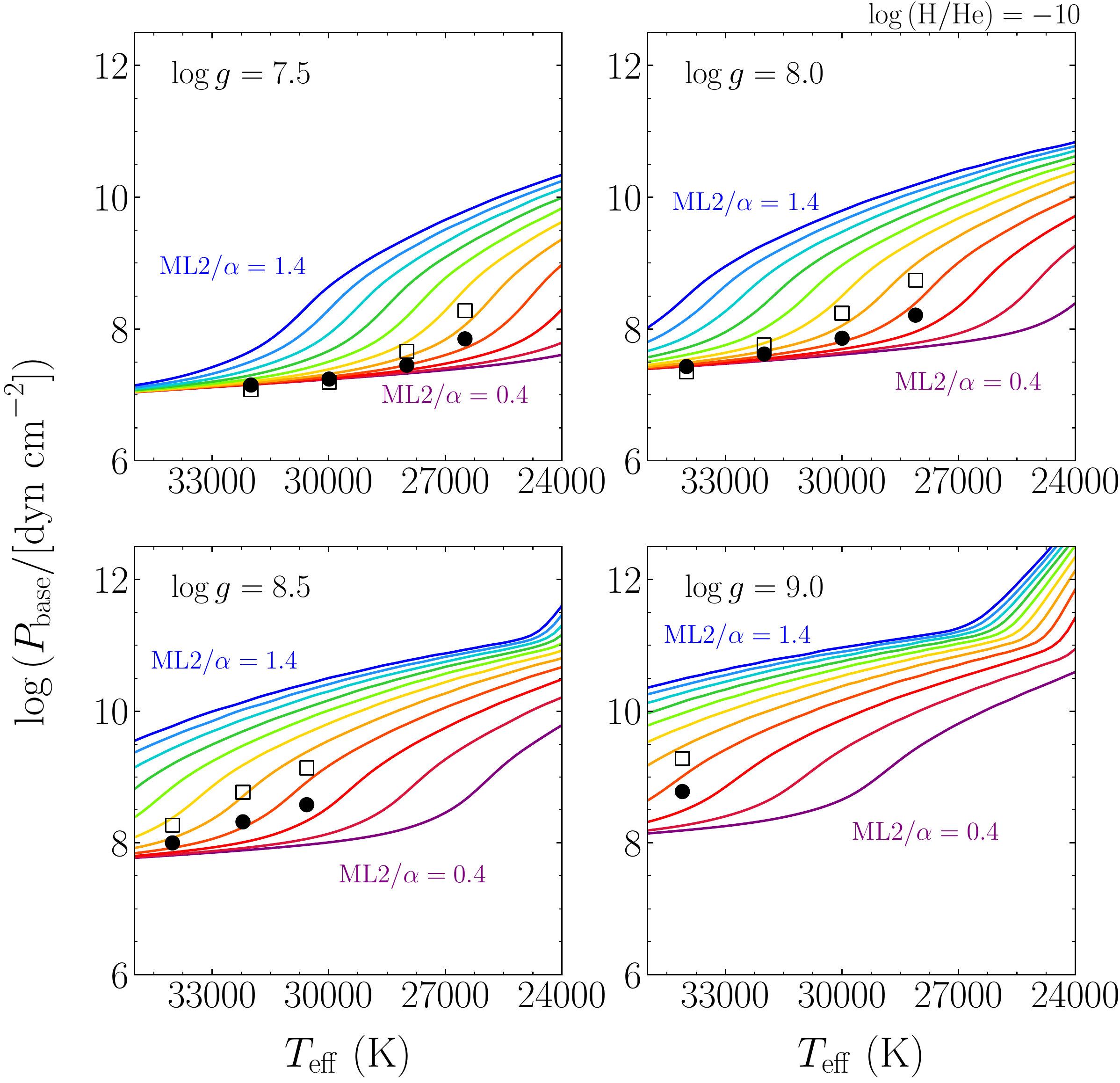}
    \caption{Similar to Fig.~\ref{fig:t_closed} but for pressure at the base of the convection zone.}
    \label{fig:p_closed}
\end{figure*}

For the closed bottom 3D models (examples shown in Figs.~\ref{fig:closed_bottom_s} and~\ref{fig:closed_bottom_hcon}) both the Schwarzschild and flux boundaries can be directly probed and the \threeD~temperature and pressure values at the two boundaries can be extracted. Similarly, from 1D envelope structures we also have access to the temperature and pressure at the bottom of the 1D \s~boundary. These quantities are displayed in Figs.~\ref{fig:t_closed} and \ref{fig:p_closed}.

For each 3D model with given atmospheric parameters, we interpolate over 1D envelopes with the same atmospheric parameters but varying values of \mlt, in order to find the \mlt~value that gives the same temperature and pressure at the base of either \s~or flux boundary of the 3D convection zone. We refer to these calibrated \mlt~values as \mlts~and \mltf~for \s~and flux boundaries, respectively. The calibrated \mlt~parameters between temperature and pressure generally agree within $\approx 0.05$ even in the most extreme cases such as \logg~=~9.0 shown in Figs.~\ref{fig:t_closed} and \ref{fig:p_closed}. Therefore, we take an average of the two \mlt~values. This gives us an indication of the average temperature gradient in the vicinity of the base of the convection zone.

A larger \mlt~value means that the convection zone extends deeper into the envelope and thus both the temperature and pressure are larger at the base. As \teff~increases for \logg~$=7.5$~and~8.0 models, the different \mlt~envelopes start to converge, yet we can still deduce that the calibrated \mlt~value in this \teff~range must be on the lower end of our \mlt~range, meaning that the convective efficiency is very low. 

The blue edge of the DBV instability strip is thought to be related to recombination of the main constituent of the atmosphere, which also causes convection to set in. Our 3D models indicate that a lower \mlt~value than 1.25 (the value used by \cite{vangrootel2017} to determine the theoretical blue edge) best represents the base of the convection zone both for \s~and flux boundaries. In general, with the lowering of \mlt~value, convection will occur later in the white dwarf's evolution (i.e. at lower \teff). The theoretical location of the blue edge of the instability strip should therefore be at a lower \teff~than predicted by current studies. 

With closed bottom models we can also directly calculate \q~for either convection zone boundary. In Figs.~\ref{fig:q_teff} and~\ref{fig:q_teff_y-2} we compare 3D \q~to the predictions of 1D envelopes. Unlike the DA case \citep{tremblay2015} we do not find that mass-calibrated \mlt~values are similar to the temperature- and pressure-calibrated \mlt~values. As the mass is calculated independently of either temperature or pressure, a disagreement is not unexpected since 1D models cannot reproduce all of the dynamic quantities of 3D models. This is clearly shown in Figs.~\ref{fig:closed_bottom_s} and ~\ref{fig:closed_bottom_hcon}, where we plot \threeD~structures and corresponding 1D atmospheric models of \cite{bergeron_db_2011} calculated at calibrated \mlts~and \mltf~values, respectively. As expected, the \threeD~and 1D structures agree in the vicinity of either boundary, but the overall 1D and \threeD~structures do not agree well. For all closed bottom models at \logg~$=$~$7.5$ and $8.0$, the masses included in the 3D convection zones diverge off the 1D envelope predictions, such that they are much smaller than what is possible to achieve in 1D within our range of \mlt~values.

In Figs.~\ref{fig:q_teff} and~\ref{fig:q_teff_y-2} flux and \s~boundary reversal is observed, just like in 3D DA models. As mentioned previously, the reversal is due to kinetic energy flux and if neglected it is not observed \citep{kupka2018,tremblay2015}. Such reversal does not occur in 1D models, as kinetic energy flux is not considered.

For studies in need of the physical conditions near the base of the convection zone, the calibrations shown in Figs~\ref{fig:t_closed} and~\ref{fig:p_closed} and listed in Tabs.~\ref{tab:3d_db_modelsclosed10} to \ref{tab:3d_modelsclosed2} of Appendix~\ref{ap:tables} should be used. The masses listed in those tables are the 1D convection zone masses found from 1D envelopes calculated at 3D calibrated \mlt~values. For studies where such approximations are not adequate, the direct use of 3D structures would be more beneficial.

\subsection{Open bottom models}

\begin{figure*}
    \centering
    \begin{subfigure}[b]{\columnwidth}
        \includegraphics[width=\columnwidth]{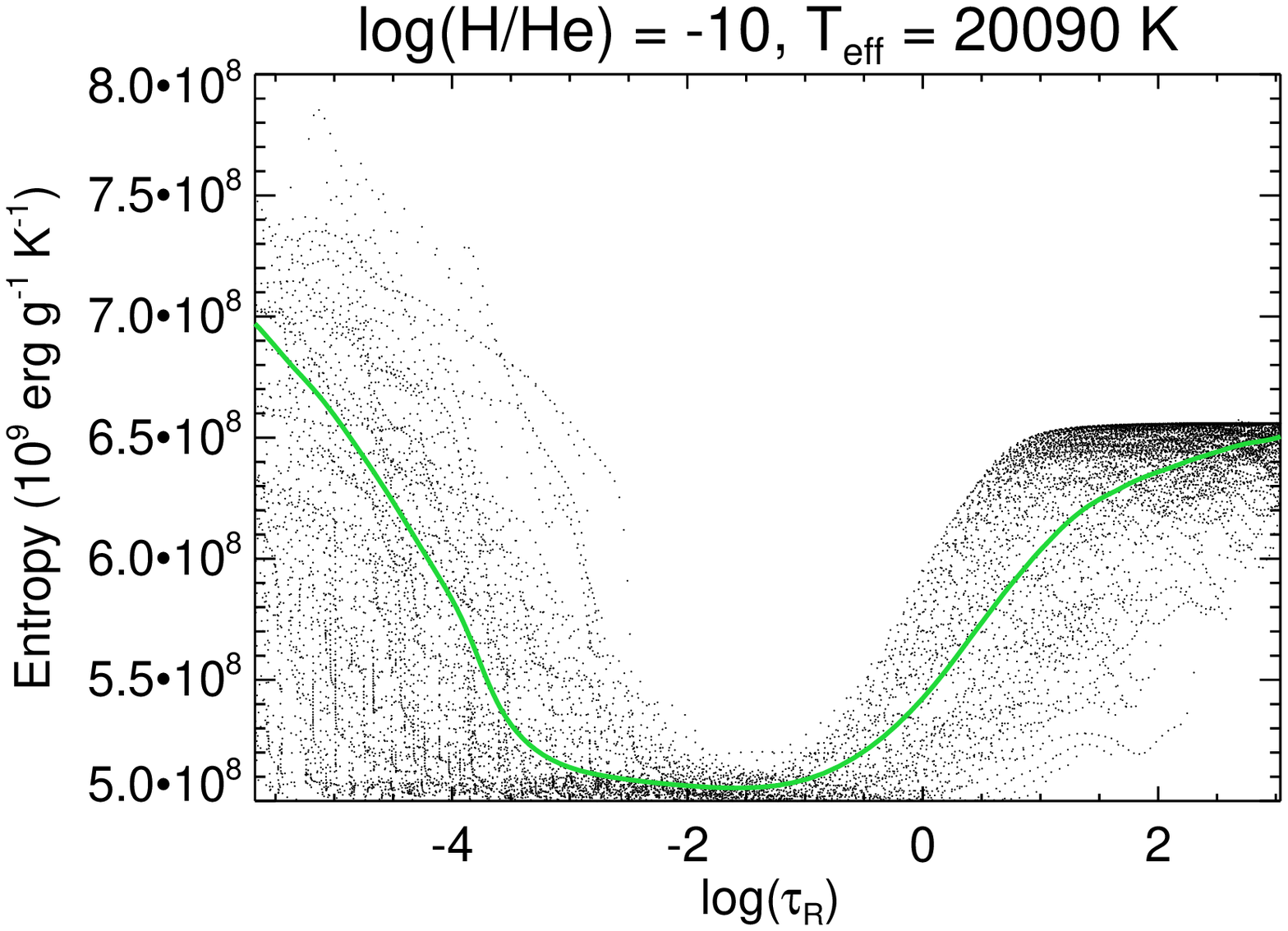}
    \end{subfigure}
    ~
    \begin{subfigure}[b]{\columnwidth}
        \includegraphics[width=\columnwidth]{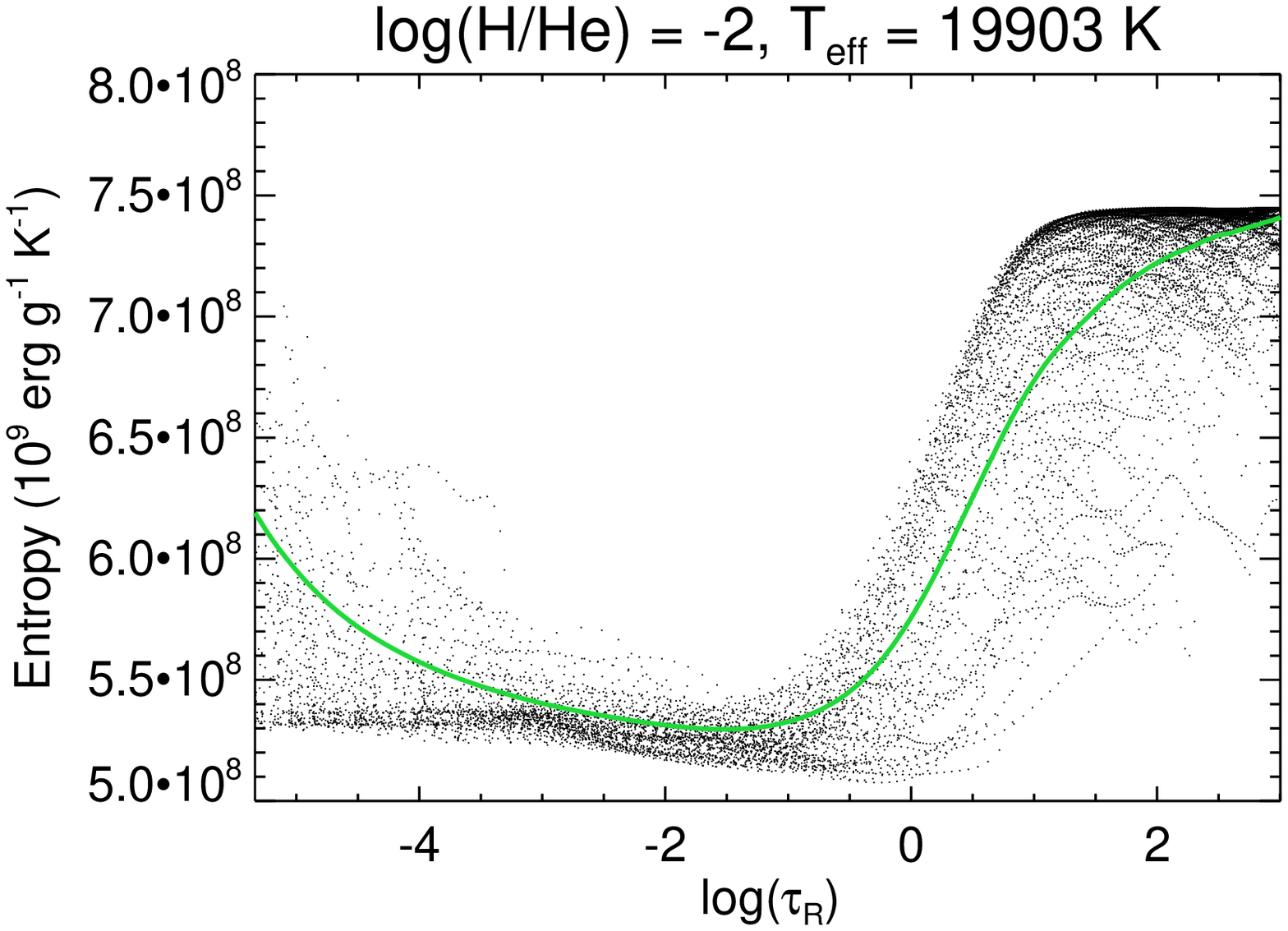}
    \end{subfigure}

    \begin{subfigure}[b]{\columnwidth}
        \includegraphics[width=\columnwidth]{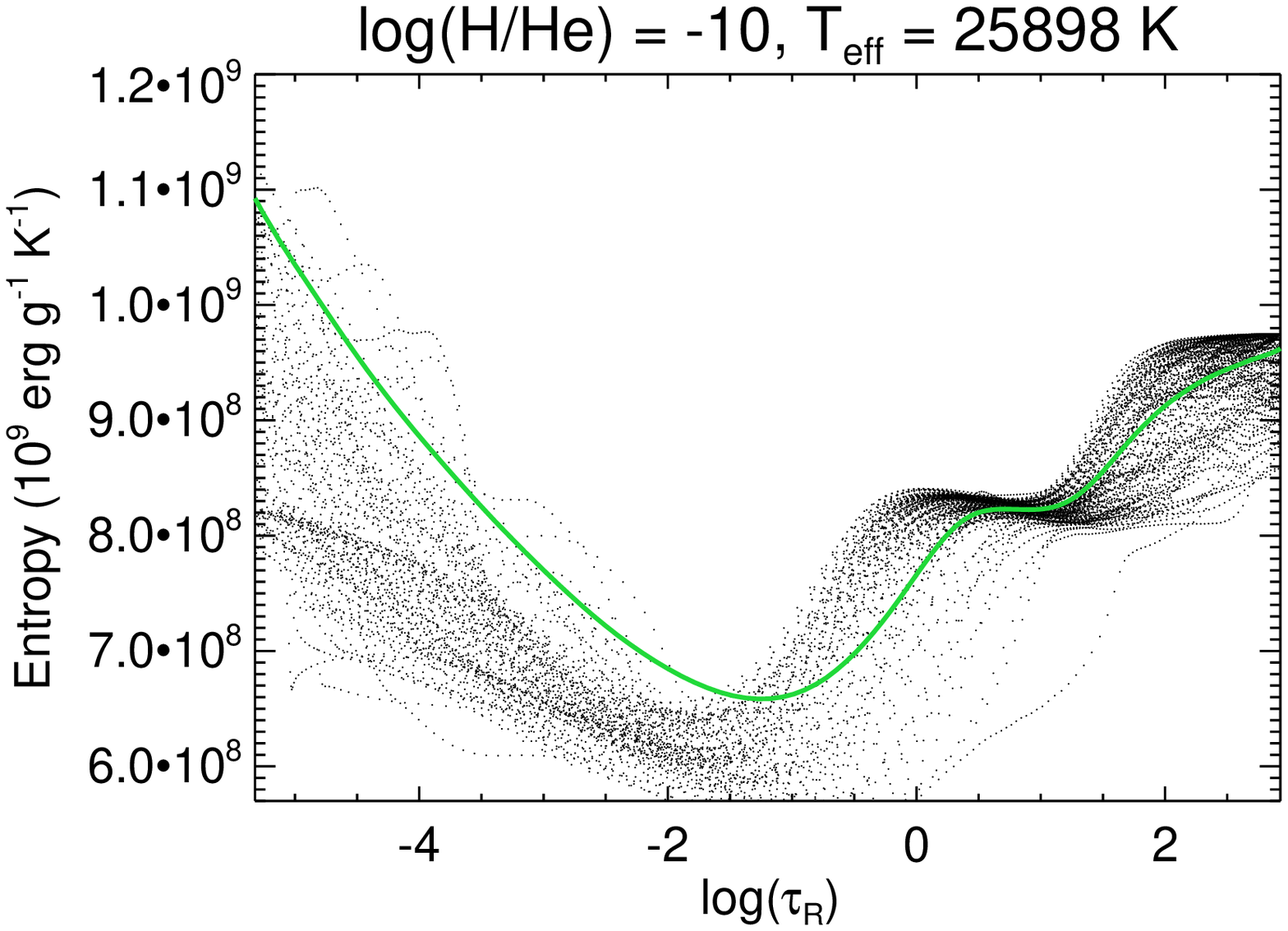}
    \end{subfigure}
    ~
    \begin{subfigure}[b]{\columnwidth}
        \includegraphics[width=\columnwidth]{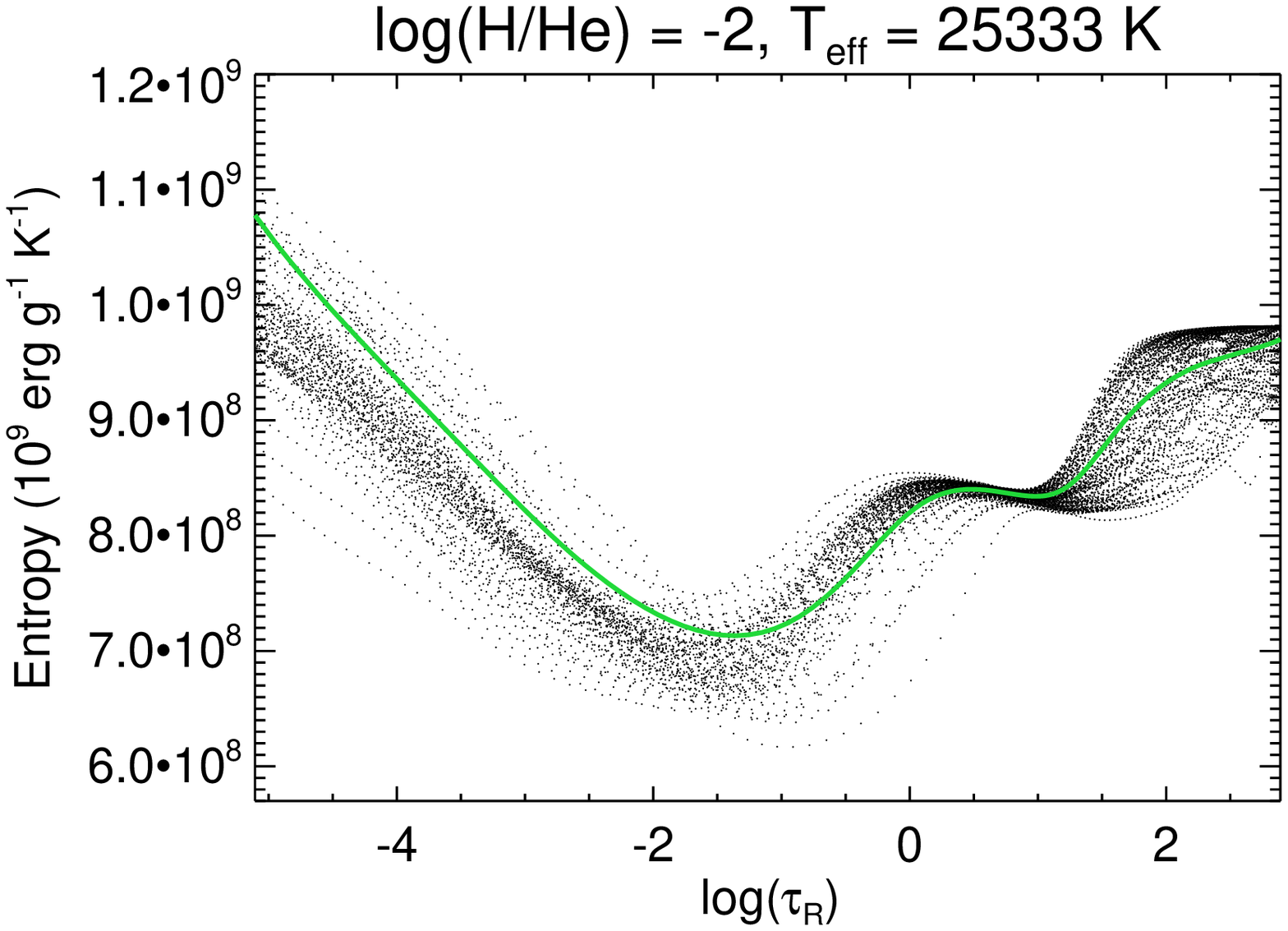}
    \end{subfigure}
    \caption{The spatially- and temporally- resolved entropy of \logg~$=8.0$ 3D open bottom models. The top two plots show the entropy stratification when only the He~I convection is present, whereas the bottom two panels show models with both He~I and He~II convection zones. In green we plot the averaged entropy over constant geometric depth and time. Although the average entropy does not reach the adiabatic value near the bottom of the simulation, it is clear that the spatially- and temporally- resolved entropy has a plateau at deeper layers, which corresponds to the inflowing entropy, an input parameter of our 3D models.}\label{fig:open_models_entropy}
\end{figure*}

For open bottom models we are unable to probe the bottom of the convection zone as our simulations are not deep enough. We can, however, exploit the fact that in 3D models a fraction of upflows from the bottom of the deep convection zone retain their adiabatic entropy almost all the way up to the observable atmospheric layers by not interacting with neighbouring downflows via heat exchange \citep{stein_nordlund1989}. This means that the spatially- and temporally-resolved entropy has a plateau corresponding to this adiabatic entropy value and it can be used to calibrate \mlt~\citep{steffen1993,ludwig1999}. Example entropy plateaus are shown in Fig.~\ref{fig:open_models_entropy} for \y~=~$-10.0$ and \y~$=-2.0$ models, where we also plot the temporally- and horizontally-averaged entropy stratifications. The averaged entropy is lower and does not reach the adiabatic entropy as it also considers the small entropy of the downflows. For \co~the adiabatic entropy value is the inflowing entropy input parameter and an entropy plateau is observed in all open bottom simulations.

For each 3D model with given atmospheric parameters, we interpolate over the different \mlt~1D envelopes with the same atmospheric parameters to find the 1D entropy at the bottom of the \s~boundary that best matches the 3D adiabatic entropy. We show this in Fig.~\ref{fig:s_teff}. The entropy of closed bottom models is also shown, but for these models we do not use the entropy to calibrate. This is because we have already calibrated \mlt~directly in Sec.~\ref{sec:closed} and generally for closed bottom models the upflows are not adiabatic in any portion of the convection zone.


The adiabatic entropy value is for the 3D Schwarzschild boundary only. We cannot access the flux boundary for open bottom models. Instead, we use the results from closed bottom models to estimate the \mlt~value that best represents the flux boundary for open bottom models. For closed bottom models that do not show the flux and \s~boundary reversal we find the relation \mltf $= 1.17$ \mlts~with a standard deviation of around 3\%. A similar result of \mltf $= 1.16$ \mlts~with a standard deviation of around 3\% was found for 3D DA models \citep{tremblay2015}. 

In Figs.~\ref{fig:q_teff} and~\ref{fig:q_teff_y-2} we show the \q~value for both open and closed bottom models with \y~$=-10.0$ and $-2.0$, respectively. Unlike the closed bottom case, we cannot directly access the bottom of either convection zone boundary for open bottom models. Thus, the masses for open bottom 3D models are extracted from the 1D envelopes with \mlt~value that best matches the 3D adiabatic entropy.

As mentioned earlier and shown in Fig.~\ref{fig:s_teff}, at the lowest \teff~the different \mlt~value envelopes converge to the same solution as convection becomes adiabatic and insensitive to \mlt~even in the upper atmosphere. In these cases, the derived mass fraction does not change significantly between the different values of the \mlt~parameter. Therefore, we propose not to interpolate for the best matching mixing length parameter, but to set it to 1.0 for both Schwarzschild and flux boundaries. 

\section{Discussion}~\label{sec:disc}

\begin{figure*}
	\includegraphics[width=1.5\columnwidth]{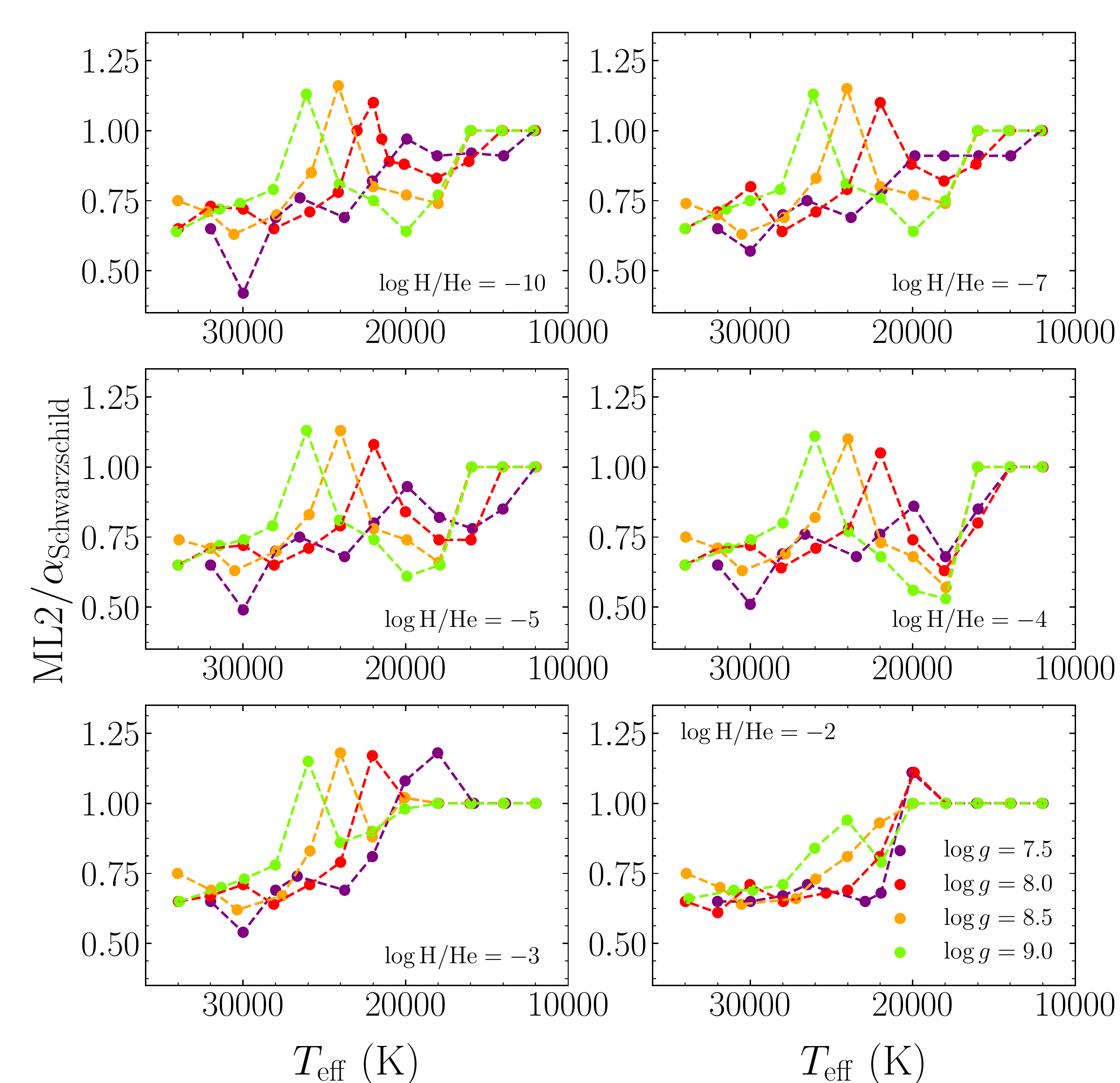}
    \caption{The calibrated mixing length parameter based on the Schwarzschild boundary is plotted as solid colour points which are connected for clarity for the same surface gravity. The value of \y~is indicated on each panel.}
    \label{fig:ml2_alpha_teff_s}
\end{figure*}

The calibrated \mlt~values are shown in Figs.~\ref{fig:ml2_alpha_teff_s} and~\ref{fig:ml2_alpha_teff_f} for the \s~and flux boundaries, respectively, and in the Appendix~\ref{ap:tables} of the Supplementary Material. In all cases, \mlt~values are smaller than what is often used in evolutionary models, i.e. \mlt~$=1.25$. This means that 3D models predict lower convective efficiencies. Given that the value of 1.25 is based on matching observed and model spectra and therefore describes the convective efficiency in the photosphere, it is not unexpected that it is different to the convective efficiency at the bottom of the convection zone. Interestingly, the mean convective efficiency for DB/DBA white dwarfs is very similar, or only slightly larger, to that of DA stars \citep{tremblay2015}. 

The plateaus observed at low \teff~are artificial. They are the consequence of fixing the value of \mlts~$=$~\mltf$=1.0$ for \teff~where the structures become insensitive to the \mlt~parameter. A similar effect can be observed at the highest \teff, where the calibration is forced to values of 0.65 for both \mlts~and \mltf, as none of the 1D \mlt~values can reproduce the boundaries of the 3D convection zone. Since the convective zone is in any case very small and inefficient in this regime, the fixed value may not be a concern for some applications. If on the other hand detailed convective properties are required, it is more appropriate to directly use 3D models which also include velocity overshoot (see Sect.~\ref{sec:v}). 

\begin{figure*}
	\includegraphics[width=1.5\columnwidth]{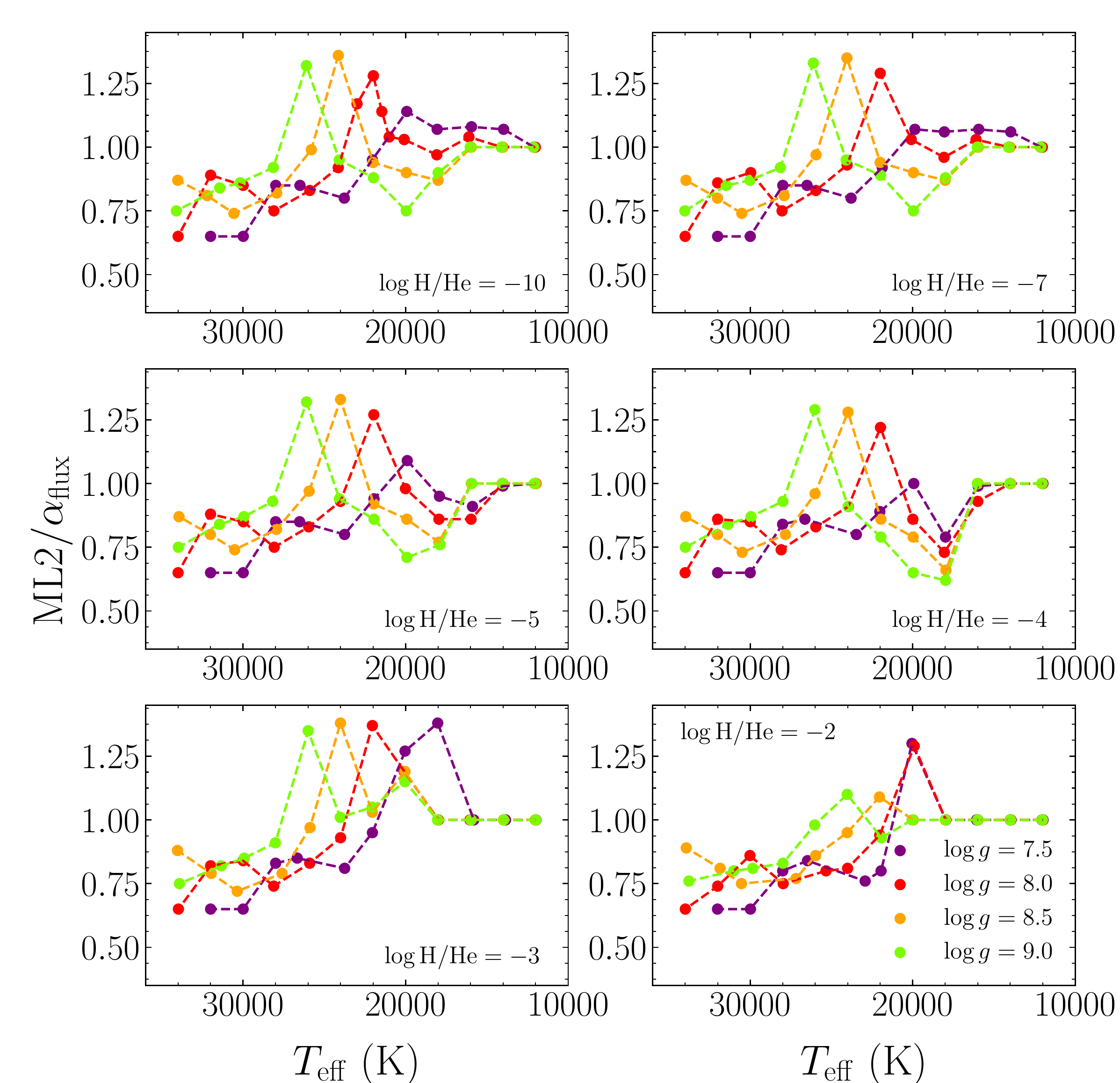}
    \caption{Same as Fig.~\ref{fig:ml2_alpha_teff_s} but for the flux boundary.}
    \label{fig:ml2_alpha_teff_f}
\end{figure*}

The peaks observed in Figs.~\ref{fig:ml2_alpha_teff_s} and~\ref{fig:ml2_alpha_teff_f} which seem to shift to higher \teff~for higher \logg, are associated with the knee-like feature of the 1D envelopes seen in Figs.~\ref{fig:s_teff}, \ref{fig:q_teff} and \ref{fig:q_teff_y-2}, which we suggest is related to the disappearance of the He~II convection zone as the white dwarf evolves to lower \teff. This transition is different in 3D, potentially because of the non-local coupling of the two convection zones. The knee-feature also means that \mlt~calibration is more sensitive in that region. 

\subsection{Calibration of the entropy jump}~\label{sec:sjump}

Studies such as \cite{magic2015} have also performed \mlt~calibrations for solar-like stars based on the entropy jump associated with superadiabatic convection. Examples of such entropy jumps can be seen in Figs.~\ref{fig:closed_bottom_s} and~\ref{fig:open_models_entropy} for closed and open bottom models, respectively. In their calibration, \cite{magic2015} define the jump as the difference between the constant entropy value of the adiabatic convection zone and the entropy minimum for both 1D and 3D models. We use a similar method to investigate more clearly the variations of \mlt~as a function of \teff.

To perform the calibration we do not use the evolutionary models presented in Sec.~\ref{sec:oneDmodels}. Instead, we use the 1D atmospheric models of \cite{bergeron_db_2011}. This grid of models spans the same range of atmospheric parameters as our 3D and 1D envelope grids, but also \mlt~values in the range $0.5 \leq$~\mlt~$\leq 1.5$ in steps of 0.25. We define the entropy jump,  $s_{\rm{jump}}$, as 
\begin{equation}
s_{\rm{jump}} = s(\log{\tau_{\rm{R}}} = 2) - s_{\rm{min}},
\end{equation}
where $s(\log{\tau_{\rm{R}}} = 2)$ is the entropy at \taur~$=2$ and $s_{\rm{min}}$ is the minimum entropy value. In the 3D case, the entropy stratification is temporally- and spatially-averaged, with the spatial average being performed over constant geometric height as before. We calculate $s_{\rm{jump}}$ both for the 3D atmospheric models, and for 1D atmospheric models calculated at different values of \mlt. We then find the value of \mlt, which we refer to as \mltjump, that best represents the given \threeD~entropy jump. In late-type stars, the entropy jump was found to decrease for increasing values of \mlt~\citep{magic2015}. This is because as convection becomes more efficient, smaller temperature gradients in the superadiabatic layers are needed to transport the same flux \citep{sonoi_mlt_evolution}. This relation holds for DB and DBA 1D models where the entropy minimum is located at the top of the He~I convection zone (see Fig.~\ref{fig:open_models_entropy} for example). It breaks down when the He~I convection zone disappears or when the entropy minimum moves to the top of the He~II convection zone. This happens for the majority of 3D closed bottom models, and therefore we only perform \mltjump~calibration for 3D open bottom models. 

\begin{figure}
	\includegraphics[width=\columnwidth]{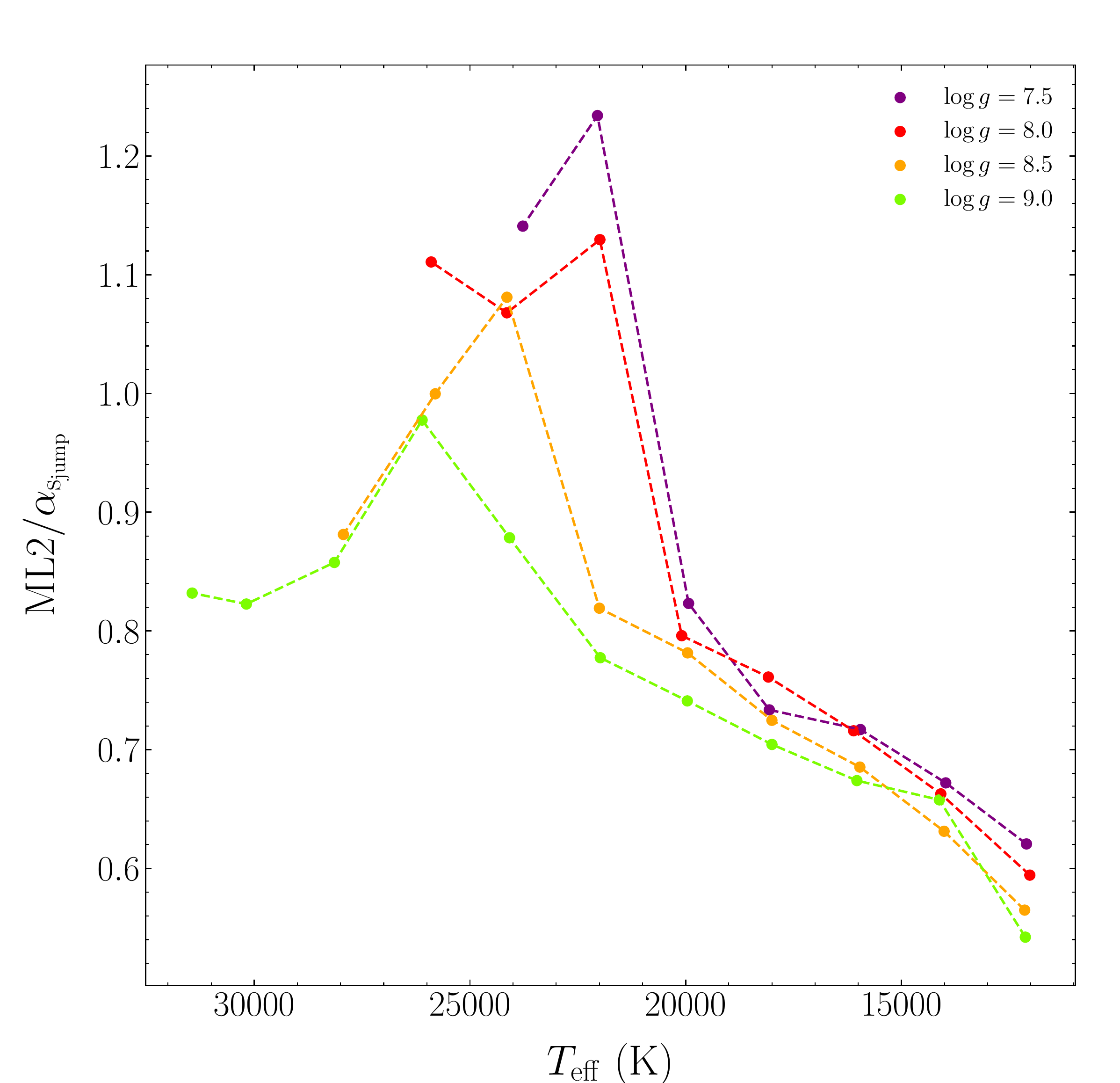}
    \caption{The calibrated mixing length parameter based on the entropy jump for open bottom 3D DB models. The solid colour points represent the \mltjump~values and are connected based on their \logg~for clarity.}
    \label{fig:sjump}
\end{figure}

We show the \mltjump~values for DB white dwarfs in Fig.~\ref{fig:sjump}. Similar results were found for DBA white dwarfs. For all \logg~apart from 7.5, the peaks observed in \mltjump~are at the same \teff~as the peaks observed for \mlts~and \mltf. By looking at the structures directly, the peaks are clearly associated with the disappearance of the second-hump in the entropy profile due to He~II convection zone as the white dwarf cools to lower \teff. Examples of double peaked entropy profiles are shown in Fig.~\ref{fig:open_models_entropy}. 

For atmospheric parameters where convection is sensitive to the \mlt~value (e.g. the calibrated value of \mlt~is not fixed in Figs.~\ref{fig:ml2_alpha_teff_s} and~\ref{fig:ml2_alpha_teff_f}), we find reasonable agreement between the \mltjump, \mlts~and \mltf~calibrations.

\cite{magic2015} found that their \mlt~values based on the entropy jump were higher than the \mlt~values based on the adiabatic entropy (\mlts). They attribute this to the 1D entropy minimum being lower than the \threeD~entropy minimum, which is also the case for our lower \teff~models. This explains why at low \teff~we find \mlts~and \mltf~values that are larger than the value of \mltjump~(for example, \teff~$\lesssim 20\,000$ K for \logg~$=8.0$ DB models).

From the studies of \mltjump, \mlts~and \mltf~it is apparent that the peaks in \mlt~values are observed close to the red edge of the DBV instability region. This means that in terms of the 3D picture, the mixing length changes quite rapidly in the region where pulsations are empirically observed to stop. As current DBV studies use an \mlt~value of 1.25, and the peak is closer to this value than the calibrated \mlt~values at other \teff, we expect that our calibration will not significantly alter the current theoretical DBV studies at the red edge of the instability strip. 

\subsection{Calibration of the maximum convective flux}~\label{sec:fmax}

\begin{figure}
	\includegraphics[width=\columnwidth]{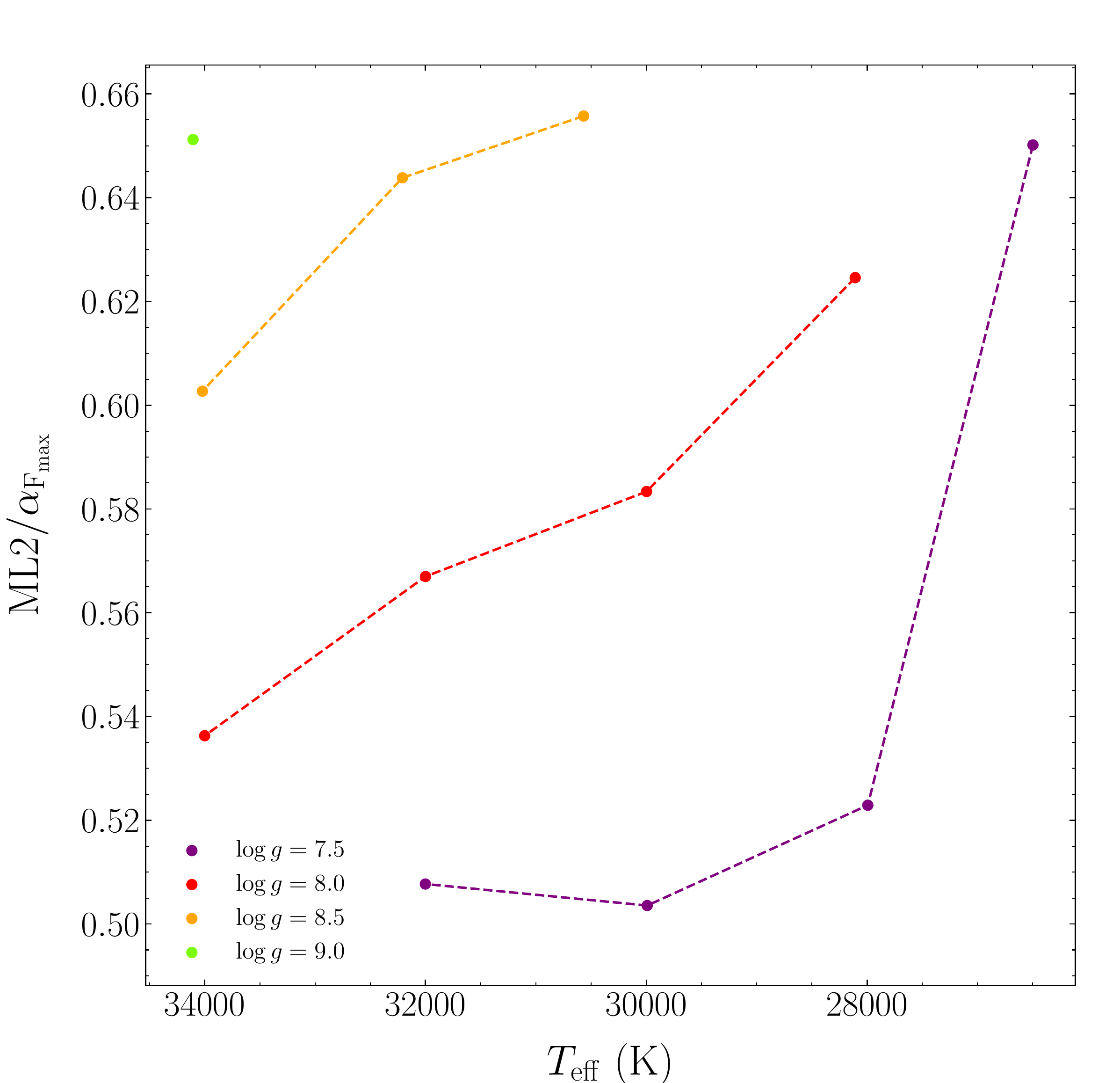}
    \caption{Same as Fig.~\ref{fig:sjump}, but for \mlt~calibration based on the maximum convective flux for 3D closed bottom models.}
    \label{fig:fmax}
\end{figure}

An alternative way to calibrate the \mlt~values for closed bottom models has been proposed by \cite{tremblay2015}. The calibration is based on the maximum value of the convective-to-total flux. This better represents the total amount of energy transported by convection as shown for DA white dwarfs by \cite{tremblay2015}. We perform this calibration for DB and DBA closed bottom models using the 1D atmospheric models of \cite{bergeron_db_2011}, i.e. same grid that was used in Sec.~\ref{sec:sjump}, but with additional grids at \mlt~$=0.55$, 0.60, 0.65 and 0.70 as convective flux changes significantly with \mlt~value. Our results are shown in Fig.~\ref{fig:fmax}. In Fig.~\ref{fig:closed_bottom_hcon}, we confirm that \mltfmax~calibration does indeed better reproduce the overall shape of DB (and DBA, although not shown) convection zones.

Overall, the \mltfmax~results are similar to \mlts~and \mltf~calibration. We find inefficient convection resulting in small convection zones. \cite{montgomery2004} performed an equivalent calibration of maximum convective flux using their 1D non-local envelope models of DB white dwarfs. They found \mlt~$\approx 0.5$ for \logg~$=8.0$, 28\,000~K~$\leq$~\teff~$\leq$~33\,000~K DB models, whereas we find 0.64~$\gtrsim$~\mlt~$\gtrsim$~0.5 for the same atmospheric parameter range. Both studies therefore suggest that convection is less efficient than what is currently assumed. When comparing DA and DB white dwarfs in the regime of very inefficient convection (closed bottom models in our case), \cite{montgomery2004} found that for given $F_{\rm{convective}}/F_{\rm{total}}$, DB stars have lower values of \mltfmax, but larger convection zone sizes. They attribute this to the He~II convection zone being deeper than the H~I counterpart, allowing the same amount of convective flux to be transported more efficiently and therefore with a smaller value of \mlt. Comparing our results to the 3D DA calibration of \cite{tremblay2015}, we also find that DB white dwarfs have smaller \mltfmax~values and larger convection zone sizes, in agreement with \cite{montgomery2004} results.

\subsection{Calibration of velocities}~\label{sec:v}

\begin{figure}
	\includegraphics[width=\columnwidth]{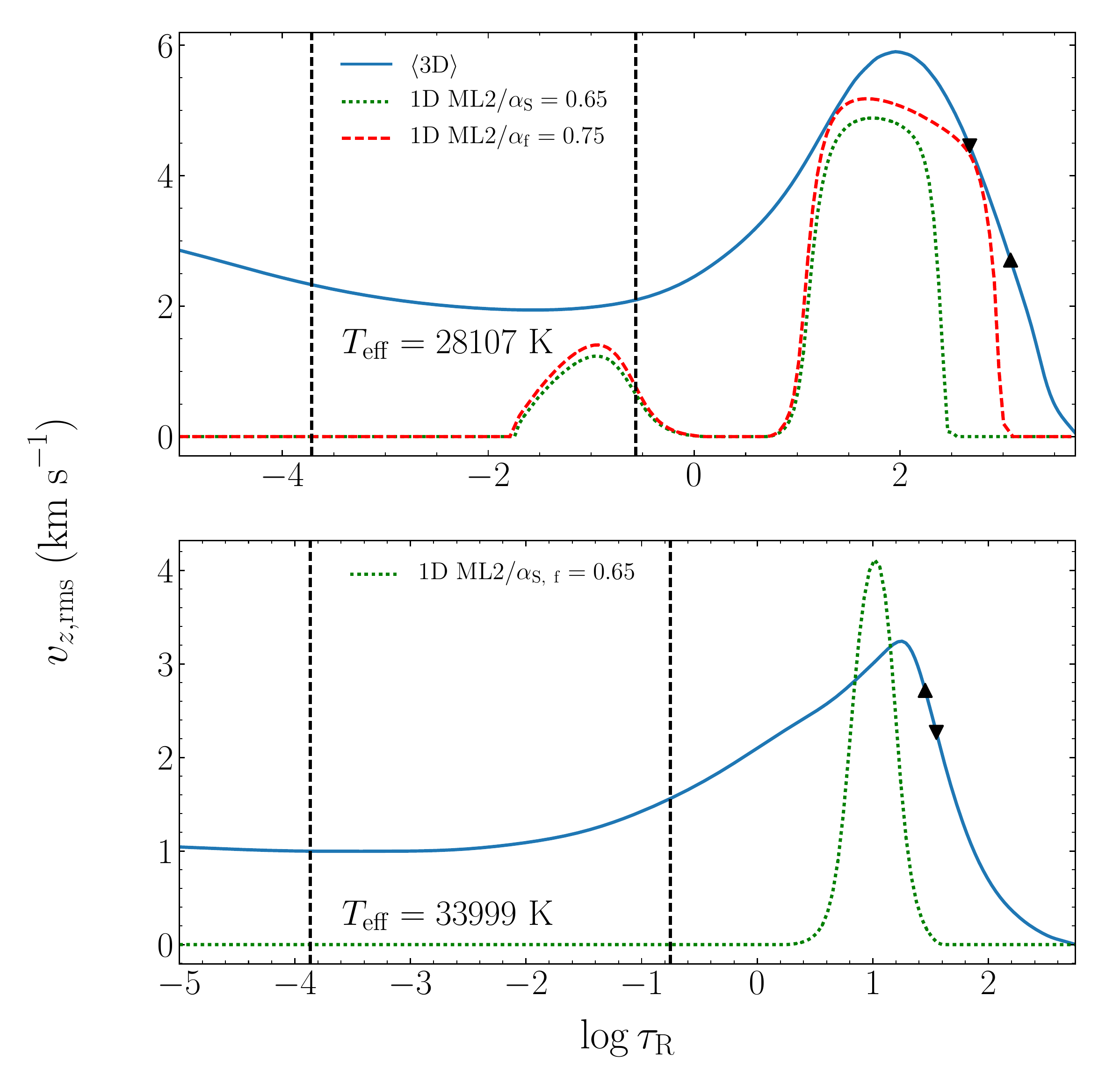}
    \caption{The vertical root mean square velocity as a function of \taur~at two different \teff~for \logg~$=8.0$ DB models. The \threeD~\vrms~is shown in solid blue. The 1D models with \mlt~$=$ \mlts~and \mltf~are shown as dotted green and red dashed lines, respectively. The bottom of the \s~and flux boundaries are shown as downward- and upward-pointing triangles. The dashed black lines indicate the top and bottom of the optical light forming region. The 1D structures are unable to reproduce \threeD~velocities especially outside the convective regions. In the upper layers (\taur < -3), the \threeD~convective velocities have an important contribution from waves in the simulation.}
    \label{fig:v}
\end{figure}

Unlike in 1D models, in 3D simulations we expect there to be significant macroscopic diffusion at the bottom of the convection zone caused by momenta of downflows. We refer to this region as the velocity overshoot region, which overlaps with the flux overshoot region shown in Fig.~\ref{fig:closed_bottom_hcon} where negative flux is found. The velocity overshoot both includes and extends beyond the flux overshoot region. The overshoot region can be thought of as an extension to the more traditional convection zones discussed in this paper, especially for studies of metal diffusion in the atmospheres of white dwarfs. If included, it would mean larger convection zones than presented in this paper. In Fig.~\ref{fig:v} we compare the velocities of our \threeD~and 1D structures. In 1D the convective velocities are only non-zero inside the \s~convection zone, whereas in 3D, the velocities are significant even beyond the \s~and flux boundaries. As long as these convective velocities result in a macroscopic diffusion process that is more efficient than microscopic diffusion, metals are expected to be fully mixed in the convection zone rather than diffuse out of it. Convective overshoot could also significantly enhance the dredge-up of carbon from the interior \citep{dufour2005} if the size of the superficial helium layer is small enough to allow convection to reach the underlying carbon layer.

Macroscopic diffusion can only be studied in 3D models with closed bottom. Yet, it is expected that all 3D models, including those with open bottom, will have overshoot both at the bottom and top of their convection zones, due to the dynamics of the convective flows. In order to study velocity overshoot for lower \teff~at which
we currently only have open bottom models, a new grid of deep closed bottom models would have to be calculated.

\cite{cunningham2019} have recently performed an in-depth study of overshoot in 3D DA closed models, finding that the mixed masses can be as much as 3 dex larger than currently used. Such a study for 3D DB and DBA models is beyond the scope of the current paper. As such, we do not attempt to perform any \mlt~calibration based on velocities.

\subsection{Impact of metals on size of the convection zone}

In order to test the effect of metals on the size of the convection zone, we calculate two sets of 3D models with and without metals at two selected \teff~values. We use the 1D atmospheric code of \cite{koester2010} to calculate input equations of state and opacity tables. When including metals, we use the metal composition and abundances of \wdz~determined by \cite{dufour2012}, as well as their determined hydrogen abundance of \y~$=-5.73 \pm 0.17$. We base our atmospheric composition on this white dwarf because it is one of the most polluted objects with 14 elements heavier than helium present in its atmosphere. Our aim is not to replicate exactly the atmospheric parameters determined by \cite{dufour2012} but rather to study the effect of strong metal pollution on 3D models.

We start our models from two computed simulations of the 3D DBA grid with \y~$=-5.0$, \logg~$=8.0$ and \teff~$\approx 14\,000$~K and $\approx 20\,000$~K. As \y~is ultimately controlled by the input tables, the \y~value of the starting model does not matter, but for convergence it is desirable to start with the closest available hydrogen abundance. Although, a value of \logg~$=8.4 \pm 0.2$ was determined by \cite{dufour2012}, we instead use \logg~$=8.0$, more in line with the recent determination of \logg~$=8.05 \pm 0.15$ by \cite{gentilefusillo2019,gentilefusillo2019cat}. 

As \teff~is only recovered after the model is run, for each set of models we tried to achieve an agreement of around 100 K between the models with and without metals. We find that including our selected metal-rich composition in a 3D model decreases the \teff~by around $1\,500$ K given the specified inflowing entropy at the bottom boundary (using the same entropy zero point). For example, the non-metal \teff~value of one model is $13\,975$~K, whereas the \teff~of the metal version is $12\,497$~K with the same physical conditions at the bottom. In order to get an agreement of $\approx 100$ K between models with and without metals, we had to increase the entropy of the inflowing material at the bottom boundary. From Figs.~\ref{fig:s_teff} and~\ref{fig:q_teff} it is clear that higher inflowing entropy means smaller convection zone. Therefore, we can speculate that with the inclusion of metals, the size of the convection zone becomes smaller for the same \teff. This is not unexpected, since similarly to hydrogen, metals increase the total opacity. 

\begin{figure}
	\includegraphics[width=\columnwidth]{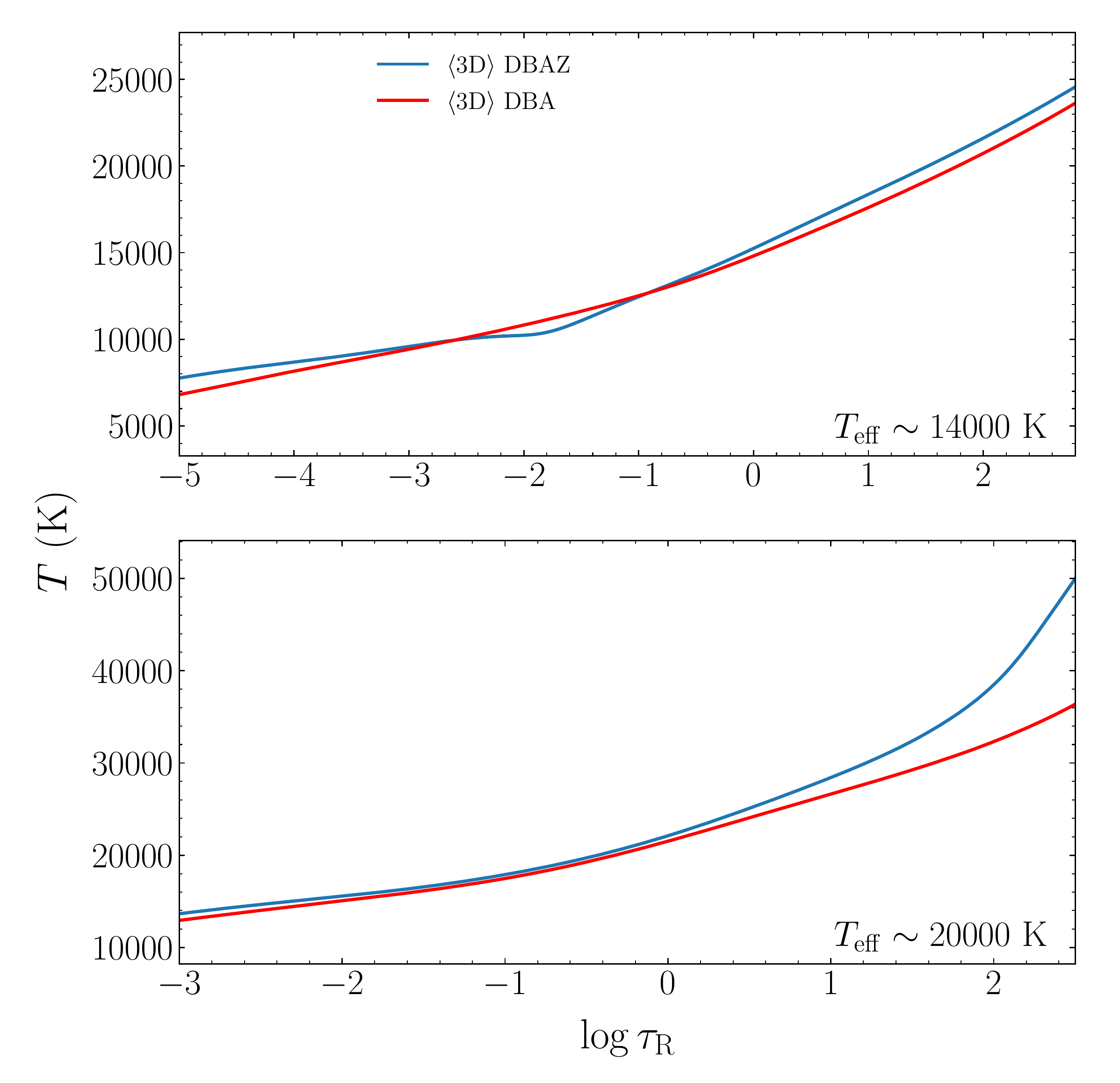}
    \caption{Temperature stratification of 3D models with and without metals at two different \teff~values. The \threeD~structures for 3D DBAZ models are shown in solid blue, whereas the non-metal 3D models are in plotted in solid red.}
    \label{fig:dbz_temp}
\end{figure}

To find the mass of the convection zone we utilise the envelope code described in \cite{koester_kepler_2015} with our calibrated \mlt~parameter. The code takes the last point in a given \threeD~atmospheric structure as a starting point for calculating the corresponding envelope. The envelope code is 1D and therefore depends on the mixing length theory. As per our calibration based on \y~$=-5.0$, \logg~$=8.0$ 3D models, we use \mlt~$=1.0$ and 0.80 for \teff~$\approx 14\,000$ K and $20\,000$ K models, respectively. We do not perform any additional mixing length parameter calibration beyond what has been described in previous sections. The total mass of the white dwarf is assumed to be 0.59\msun~with radius of 0.0127$R_{\odot}$. The \cite{saumon1995} equation of state is used and only hydrogen and helium atoms are considered. Metals are ignored as they do not impact the envelope structure as long as they are a trace species. Therefore, the difference in the mass of the convection zone between the metal and non-metal models arises from the fact that the 3D atmospheric structures are different (see Fig.~\ref{fig:dbz_temp}). In Tab.~\ref{tab:dbaz} we show the change in the mass of the convection zone with the addition of metals. We find that in the \teff~$\approx 14\,000$~K case, the mass of the convection zone decreases by a factor of 2 (or 0.31 dex) when metals are included. For the \teff~$\approx 20\,000$~K case, a similar change of 0.45 dex is observed. In both cases it would mean that for the same metal abundance observed, the total mass of metals present would be smaller using the appropriate metal-rich model atmosphere. For \teff~$\approx 14\,000$ K, the change in the mass of the convection zone with the inclusion of metals can be mimicked by increasing the hydrogen abundance from \y~$=-5.0$ to $-3.0$. Similarly, at \teff~$\approx 20\,000$ K, the increase of \y~from $-5.0$ to somewhere between $-3.0$ and $-2.0$ gives a change in mass similar to the effect of metals.

In terms of the 3D picture, the effect of metals on the size of the convection zone is moderate, especially since \wdz~is one of the most heavily polluted white dwarfs. However, the effect of metals on spectroscopic 3D corrections for \teff~and \logg~are still to be explored. Fig.~\ref{fig:dbz_temp} suggests that changes in the structure of the light forming layers are important especially at lower Teff. 

\newpage

\begin{table}
	\centering
	\caption{Change in the convection zone mass from addition of metals (DBAZ) in a helium-rich DBA white dwarf. The DBAZ models use the metal abundances of \wdz~determined by \protect\cite{dufour2012}.}
	\label{tab:dbaz}
	\begin{tabular}{lccccr} 
		\hline
		$\log{g}$ & $T_{\rm{eff}}$ & Change in convection \\
		& (K) & zone mass (dex) \\ 
		\hline
		8.0 & $\approx 14\,000$ K & $-$0.31 \\
		8.0 & $\approx 20\,000$ K & $-$0.45 \\
        \hline
	\end{tabular}
\end{table}

\newpage

\section{Summary}~\label{sec:sum}

With 285 3D \co~atmospheric models of DB and DBA white dwarfs, we have calibrated the mixing length parameter for the use of 1D envelope and evolutionary models. Our results are applicable for studies in need of convection zone sizes, for example for asteroseismological and remnant planetary systems analyses. 

As the nature of the convection zone boundaries is more complex in 3D than in 1D, two definitions of the boundary were used for calibration, the \s~and flux boundaries. Overall, values of both \mlts~or \mltf~are lower than what is typically used in envelope and evolutionary models, meaning that convection is less efficient in 3D models. On average, for \logg~$= 8.0$ models with 18\,000~K~$\lesssim$ \teff~$\lesssim$~30\,000~K, we find \mlts~$\approx 0.80$ and \mltf~$\approx 0.9$. This is similar to \mlt~parameters calibrated for 3D DA white dwarfs \citep{tremblay2015}. 

Near the blue edge of the DBV instability strip, we find that the calibrated \mlt~values are much lower than the value of 1.25 recently used in the theoretical seismological study of \cite{vangrootel2017}. Therefore, in 3D, efficient convective energy transport sets in at a lower \teff. As the set-in of significant energy transport by convection is related to the blue edge of the strip, the 3D results would potentially mean lower \teff~of the theoretical blue edge. Note that compared to the empirical blue edge of \teff~$\approx$~31\,000~K at \logg~$\approx$~7.8 \citep{shipman2002,provencal2003,hermes2017,myprecious}, the current 1D theoretical blue edge of \teff~$\approx$~29\,000~K at \logg~$\approx$~7.8 is already too low in comparison (see Fig. 4 of \citealt{vangrootel2017}). 

In terms of determining the \teff~and \logg~values from spectroscopy, we recommend using \mlt~$=1.25$ (but see \citealt{myprecious} for details of 3D DB corrections). However, it is clear that the actual efficiency of convection in the atmosphere has little to do with the \mlt~$=1.25$ value calibrated from spectroscopic observations.

The current evolutionary models of white dwarfs can be improved by including our \mlt~calibrated values. 3D models also provide the best available estimate for the masses of convection zones of DB and DBA white dwarfs which are relevant for studies of remnant planetary systems. We illustrate this by calculating example 3D DBAZ models. However, our calibration does not consider velocity overshoot which could increase the mixing mass by orders of magnitude. In most of the models presented here, however, we cannot currently do any overshoot studies as the convection zones are too large to model. For the select few models at the highest \teff~of our grid, the overshoot region can be directly accessed and could be used for direct investigation, similar to what has been achieved for DA white dwarfs \citep{cunningham2019}.

Convection is not expected to have any direct impact of the derived ages of white dwarfs, up until the convection zone grows large enough to reach the core, directly coupling the degenerate core to the surface \citep{tremblay2015}. This occurs at \teff~$ \sim $ 5\,000 K for DA white dwarfs \citep{tassoul1990, tremblay2015} and $ \sim $ 10\,000 K for DB white dwarfs \citep{tassoul1990, macdonald1991}. However, at these \teff~convection is adiabatic and therefore loses its sensitivity to the \mlt~parameter. Therefore, we do not expect our calibration of the \mlt~parameter to have any direct impact on the ages derived from evolutionary models. However, the 3D models can have an indirect effect on age determinations due to 3D spectroscopic corrections for \logg~and \teff~\citep{myprecious}. 3D DBA spectroscopic corrections will be derived in a future work. 

\newpage

\section*{Acknowledgements}
We would like to thank the anonymous referee for their helpful comments. This project has received funding from the European Research Council (ERC) under the European Union's Horizon 2020 research and innovation programme (grant agreement No 677706 - WD3D). B.F. has been supported by the Swedish Research Council (Vetenskapsr{\aa}det). H.G.L. acknowledges financial support by the Sonderforschungsbereich SFB\,881 ``The Milky Way System'' (subprojects A4) of the German Research Foundation (DFG).




\newpage

\bibliographystyle{mnras}
\bibliography{aamnem99,aabib} 

\newpage




\bsp	
\clearpage
\appendix
\section{Additional information}~\label{ap:tables}

Tabs.~\ref{tab:3d_models10} to \ref{tab:3d_models2} list some basic parameters of the 3D simulations. This includes the surface gravity of a given simulation, its effective temperature, the size of the box the simulation was run in, the run time and the relative bolometric intensity contrast averaged over space and time. 

Tabs.~\ref{tab:3d_db_open_models10} to \ref{tab:3d_db_open_models2} list the parameters needed for the mixing length calibration of 3D open bottom models, as well as the results of the calibration. For each 3D simulation, its surface gravity, effective temperature and the adiabatic entropy used for \mlts~calibration is included. Also given are the \mlts, \q, $T$ and $P$ values for the \s~boundary. \q, temperature and pressure are found from the 1D envelope calculated at \mlts. The same parameters are also given for the flux boundary. As the flux boundary cannot be directly accessed for open bottom models, we instead use the relation \mltf~$=1.17$~\mlts~to find \mltf.

Tabs.~\ref{tab:3d_db_modelsclosed10} to \ref{tab:3d_modelsclosed2} list the parameters needed for the calibration of the mixing length for 3D closed bottom models, as well as the results of the calibration. For each 3D simulation, its surface gravity and effective temperature are given. The mixing length calibration for closed bottom model relies on the spatially- and temporally-averaged 3D temperature and pressure at the bottom of the convection zone, and these parameters are given for both the \s~and flux boundaries. The \mlts~and \mltf~are also given, as well as the \q~for each boundary. 

\begin{table}
	\centering
	\caption{Select parameters of the 3D DB model atmospheres, where $\delta I_{\rm rms}/\langle I \rangle$ is the relative bolometric intensity contrast averaged over space and time.}
	\label{tab:3d_models10}
\begin{tabular}{lcccccr} 
		\hline
		$\log{g}$ & $T_{\rm{eff}}$ & Box size & Total run time & $\delta I_{\rm rms}/\langle I \rangle$ \\
		 & (K) & (km $\times$ km $\times$ km) & (stellar s) & (\%) \\
		\hline
        7.5 & 12098 & 1.22$\times$1.22$\times$0.58 & 33.6 & 3.6 \\
        7.5 & 13969 & 1.98$\times$1.98$\times$0.67 & 32.2 & 8.9 \\
        7.5 & 15947 & 2.86$\times$2.86$\times$1.19 & 32.2 & 16.4 \\
        7.5 & 18059 & 6.09$\times$6.09$\times$1.46 & 32.1 & 21.3 \\
        7.5 & 19931 & 11.96$\times$11.96$\times$2.39 & 34.7 & 23.4 \\
        7.5 & 22044 & 21.75$\times$21.75$\times$4.51 & 32.3 & 25.5 \\
\hline
	\end{tabular}
\end{table}

\clearpage

\begin{table}
	\centering
	\caption{Select parameters of 3D DBA model atmospheres with $\log{\rm{H}/\rm{He}} = -7$.}
	\label{tab:3d_models7}
	\begin{tabular}{lcccccr} 
		\hline
		$\log{g}$ & $T_{\rm{eff}}$ & Box size & Total run time & $\delta I_{\rm rms}/\langle I \rangle$ \\
		 & (K) & (km $\times$ km $\times$ km) & (stellar s) & (\%) \\
		\hline
        7.5 & 12098 & 1.22$\times$1.22$\times$0.58 & 32.8 & 3.6 \\
        7.5 & 13967 & 1.98$\times$1.98$\times$0.67 & 31.7 & 8.8 \\
        7.5 & 15936 & 2.86$\times$2.86$\times$1.19 & 35.0 & 16.3 \\
        7.5 & 18051 & 6.09$\times$6.09$\times$1.46 & 34.1 & 21.0 \\
        7.5 & 19865 & 11.96$\times$11.96$\times$2.44 & 32.6 & 22.3 \\
        7.5 & 21873 & 21.75$\times$21.75$\times$4.04 & 37.8 & 22.7 \\
\hline
	\end{tabular}
\end{table}

\begin{table}
	\centering
	\caption{Select parameters of 3D DBA model atmospheres with $\log{\rm{H}/\rm{He}} = -5$.}
	\label{tab:3d_models5}
	\begin{tabular}{lcccccr} 
		\hline
		$\log{g}$ & $T_{\rm{eff}}$ & Box size & Total run time & $\delta I_{\rm rms}/\langle I \rangle$ \\
		 & (K) & (km $\times$ km $\times$ km) & (stellar s) & (\%) \\
		\hline
        7.5 & 12009 & 1.22$\times$1.22$\times$0.59 & 33.1 & 3.4 \\
        7.5 & 14013 & 1.98$\times$1.98$\times$0.67 & 33.0 & 9.0 \\
        7.5 & 15886 & 2.86$\times$2.86$\times$1.19 & 33.9 & 15.7 \\
        7.5 & 17920 & 6.09$\times$6.09$\times$1.46 & 31.8 & 21.0 \\
        7.5 & 19900 & 11.96$\times$11.96$\times$2.44 & 32.2 & 23.1 \\
        7.5 & 21946 & 21.75$\times$21.75$\times$4.51 & 32.6 & 24.7 \\
\hline
	\end{tabular}
\end{table}

\newpage

\begin{table}
	\centering
	\caption{Select parameters of 3D DBA model atmospheres with $\log{\rm{H}/\rm{He}} = -4$.}
	\label{tab:3d_models4}
	\begin{tabular}{lcccccr} 
		\hline
		$\log{g}$ & $T_{\rm{eff}}$ & Box size & Total run time & $\delta I_{\rm rms}/\langle I \rangle$ \\
		 & (K) & (km $\times$ km $\times$ km) & (stellar s) & (\%) \\
		\hline
        7.5 & 11983 & 1.22$\times$1.22$\times$0.58 & 32.0 & 3.5 \\
        7.5 & 13985 & 1.98$\times$1.98$\times$0.67 & 31.8 & 9.0 \\
        7.5 & 15973 & 2.86$\times$2.86$\times$1.19 & 34.0 & 16.0 \\
        7.5 & 17979 & 6.09$\times$6.09$\times$1.46 & 34.2 & 20.4 \\
        7.5 & 19932 & 11.96$\times$11.96$\times$2.75 & 34.5 & 22.2 \\
        7.5 & 22021 & 21.75$\times$21.75$\times$4.51 & 32.2 & 23.8 \\
\hline
	\end{tabular}
\end{table}

\begin{table}
	\centering
	\caption{Select parameters of 3D DBA model atmospheres with $\log{\rm{H}/\rm{He}} = -3$.}
	\label{tab:3d_models3}
	\begin{tabular}{lcccccr} 
		\hline
		$\log{g}$ & $T_{\rm{eff}}$ & Box size & Total run time & $\delta I_{\rm rms}/\langle I \rangle$ \\
		 & (K) & (km $\times$ km $\times$ km) & (stellar s) & (\%) \\
		\hline
        7.5 & 11980 & 1.22$\times$1.22$\times$0.37 & 31.7 & 3.3 \\
        7.5 & 13855 & 1.98$\times$1.98$\times$0.67 & 36.3 & 8.8 \\
        7.5 & 15805 & 2.86$\times$2.86$\times$1.19 & 34.6 & 16.2 \\
        7.5 & 18026 & 6.09$\times$6.09$\times$1.46 & 32.1 & 21.2 \\
        7.5 & 20035 & 11.96$\times$11.96$\times$2.53 & 33.0 & 23.3 \\
        7.5 & 22043 & 21.75$\times$21.75$\times$4.51 & 31.6 & 24.8 \\
\hline
	\end{tabular}
\end{table}

\clearpage

\begin{table}
	\centering
	\caption{Select parameters of 3D DBA model atmospheres with $\log{\rm{H}/\rm{He}} = -2$.}
	\label{tab:3d_models2}
	\begin{tabular}{lcccccr} 
		\hline
		$\log{g}$ & $T_{\rm{eff}}$ & Box size & Total run time & $\delta I_{\rm rms}/\langle I \rangle$ \\
		 & (K) & (km $\times$ km $\times$ km) & (stellar s) & (\%) \\
		\hline
        7.5 & 11977 & 1.98$\times$1.98$\times$0.67 & 39.8 & 9.0 \\
        7.5 & 13995 & 2.86$\times$2.86$\times$1.19 & 38.3 & 15.0 \\
        7.5 & 16063 & 6.09$\times$6.09$\times$1.46 & 40.0 & 17.7 \\
        7.5 & 17963 & 6.09$\times$6.09$\times$1.46 & 73.4 & 20.2 \\
        7.5 & 20042 & 21.75$\times$21.75$\times$3.39 & 36.8 & 20.6 \\
        7.5 & 21944 & 21.75$\times$21.75$\times$4.59 & 33.9 & 18.2 \\
\hline
	\end{tabular}
\end{table}

\begin{table*}
	\centering
	\caption{MLT calibration for open bottom 3D DB models, where 3D~$s_{\rm{env}}$ is the 3D adiabatic entropy used for calibration, \mlts~is the calibrated \mlt~value for \s~boundary, \q$_{\rm{S}}$ is \q~for \s~boundary, $(\log{T_{\rm{b}}})_{\rm{S}}$ is the 1D calibrated temperature at the \s~boundary, $(\log{P_{\rm{b}}})_{\rm{S}}$ is the 1D calibrated pressure at the \s~boundary. The same parameters are also given for the flux boundary and are denoted by subscript `f'.}
	\label{tab:3d_db_open_models10}
	\begin{tabular}{lcccccccccr} 
		\hline
		$\log{g}$ & $T_{\rm{eff}}$ & 3D~$s_{\rm{env}}$ & \mlts & \q$_{\rm{S}}$ & $(\log{T_{\rm{b}}})_{\rm{S}}$ & $(\log{P_{\rm{b}}})_{\rm{S}}$ &  \mltf & \q$_{\rm{f}}$ & $(\log{T_{\rm{b}}})_{\rm{f}}$ & $(\log{P_{\rm{b}}})_{\rm{f}}$ \\
		& (K) & ($10^9$ erg g$^{-1}$ K$^{-1}$) & & & (K) & (dyn cm$^{-2}$) & & & (K) & (dyn cm$^{-2}$) \\
		\hline		
        7.5 & 12098 & 0.40 & 1.00 & $-$4.14 & 6.68 & 16.96 & 1.00 & $-$4.14 & 6.68 & 16.96 \\
        7.5 & 13969 & 0.44 & 0.91 & $-$4.56 & 6.63 & 16.53 & 1.07 & $-$4.54 & 6.63 & 16.55 \\
        7.5 & 15947 & 0.48 & 0.92 & $-$5.16 & 6.52 & 15.93 & 1.08 & $-$5.11 & 6.54 & 15.99 \\
        7.5 & 18059 & 0.59 & 0.91 & $-$6.57 & 6.25 & 14.51 & 1.07 & $-$6.40 & 6.28 & 14.69 \\
        7.5 & 19931 & 0.78 & 0.97 & $-$9.16 & 5.74 & 11.92 & 1.14 & $-$8.72 & 5.83 & 12.36 \\
        7.5 & 22044 & 0.94 & 0.82 & $-$11.06 & 5.42 & 10.02 & 0.95 & $-$10.82 & 5.46 & 10.26 \\
\hline
	\end{tabular}
\end{table*}

\begin{table*}
	\centering
	\caption{Same as Tab.~\ref{tab:3d_db_open_models10} but for MLT calibration of open bottom 3D DBA models with $\log{\rm{H}/\rm{He}} = -7$.}
	\label{tab:3d_db_open_models7}
	\begin{tabular}{lcccccccccr} 
		\hline
		$\log{g}$ & $T_{\rm{eff}}$ & 3D~$s_{\rm{env}}$ & \mlts & \q$_{\rm{S}}$ & $(\log{T_{\rm{b}}})_{\rm{S}}$ & $(\log{P_{\rm{b}}})_{\rm{S}}$ &  \mltf & \q$_{\rm{f}}$ & $(\log{T_{\rm{b}}})_{\rm{f}}$ & $(\log{P_{\rm{b}}})_{\rm{f}}$ \\
		& (K) & ($10^9$ erg g$^{-1}$ K$^{-1}$) & & & (K) & (dyn cm$^{-2}$) & & & (K) & (dyn cm$^{-2}$) \\		
		\hline
        7.5 & 12098 & 0.40 & 1.00 & $-$4.14 & 6.68 & 16.96 & 1.00 & $-$4.14 & 6.68 & 16.96 \\
        7.5 & 13967 & 0.44 & 0.91 & $-$4.56 & 6.63 & 16.53 & 1.06 & $-$4.54 & 6.63 & 16.56 \\
        7.5 & 15936 & 0.48 & 0.91 & $-$5.16 & 6.52 & 15.93 & 1.07 & $-$5.11 & 6.54 & 15.99 \\
        7.5 & 18051 & 0.59 & 0.91 & $-$6.57 & 6.25 & 14.51 & 1.06 & $-$6.39 & 6.29 & 14.69 \\
        7.5 & 19865 & 0.79 & 0.91 & $-$9.23 & 5.73 & 11.85 & 1.07 & $-$8.76 & 5.82 & 12.32 \\
        7.5 & 21873 & 0.94 & 0.79 & $-$11.06 & 5.41 & 10.02 & 0.92 & $-$10.83 & 5.45 & 10.25 \\
\hline
	\end{tabular}
\end{table*}

\begin{table*}
	\centering
	\caption{Same as Tab.~\ref{tab:3d_db_open_models10} but for MLT calibration of open bottom 3D DBA models with $\log{\rm{H}/\rm{He}} = -5$.}
	\label{tab:3d_db_open_models5}
	\begin{tabular}{lcccccccccr} 
		\hline
		$\log{g}$ & $T_{\rm{eff}}$ & 3D~$s_{\rm{env}}$ & \mlts & \q$_{\rm{S}}$ & $(\log{T_{\rm{b}}})_{\rm{S}}$ & $(\log{P_{\rm{b}}})_{\rm{S}}$ &  \mltf & \q$_{\rm{f}}$ & $(\log{T_{\rm{b}}})_{\rm{f}}$ & $(\log{P_{\rm{b}}})_{\rm{f}}$ \\
		& (K) & ($10^9$ erg g$^{-1}$ K$^{-1}$) & & & (K) & (dyn cm$^{-2}$) & & & (K) & (dyn cm$^{-2}$) \\	
		\hline
        7.5 & 12009 & 0.40 & 1.00 & $-$4.13 & 6.69 & 16.98 & 1.00 & $-$4.13 & 6.69 & 16.98 \\
        7.5 & 14013 & 0.44 & 0.85 & $-$4.59 & 6.62 & 16.51 & 0.99 & $-$4.56 & 6.63 & 16.54 \\
        7.5 & 15886 & 0.49 & 0.78 & $-$5.22 & 6.51 & 15.88 & 0.91 & $-$5.14 & 6.53 & 15.95 \\
        7.5 & 17920 & 0.59 & 0.82 & $-$6.58 & 6.24 & 14.50 & 0.95 & $-$6.40 & 6.28 & 14.68 \\
        7.5 & 19900 & 0.79 & 0.93 & $-$9.24 & 5.73 & 11.84 & 1.09 & $-$8.78 & 5.82 & 12.30 \\
        7.5 & 21946 & 0.94 & 0.80 & $-$11.06 & 5.42 & 10.02 & 0.94 & $-$10.83 & 5.46 & 10.25 \\
\hline
	\end{tabular}
\end{table*}

\begin{table*}
	\centering
	\caption{Same as Tab.~\ref{tab:3d_db_open_models10} but for MLT calibration of open bottom 3D DBA models with $\log{\rm{H}/\rm{He}} = -4$.}
	\label{tab:3d_db_open_models4}
	\begin{tabular}{lcccccccccr} 
		\hline
		$\log{g}$ & $T_{\rm{eff}}$ & 3D~$s_{\rm{env}}$ & \mlts & \q$_{\rm{S}}$ & $(\log{T_{\rm{b}}})_{\rm{S}}$ & $(\log{P_{\rm{b}}})_{\rm{S}}$ &  \mltf & \q$_{\rm{f}}$ & $(\log{T_{\rm{b}}})_{\rm{f}}$ & $(\log{P_{\rm{b}}})_{\rm{f}}$ \\
		& (K) & ($10^9$ erg g$^{-1}$ K$^{-1}$) & & & (K) & (dyn cm$^{-2}$) & & & (K) & (dyn cm$^{-2}$) \\	
		\hline
        7.5 & 11983 & 0.40 & 1.00 & $-$4.21 & 6.67 & 16.89 & 1.00 & $-$4.21 & 6.67 & 16.89 \\
        7.5 & 13985 & 0.44 & 1.00 & $-$4.61 & 6.62 & 16.48 & 1.00 & $-$4.61 & 6.62 & 16.48 \\
        7.5 & 15973 & 0.49 & 0.85 & $-$5.27 & 6.50 & 15.83 & 0.99 & $-$5.21 & 6.51 & 15.88 \\
        7.5 & 17979 & 0.62 & 0.68 & $-$7.01 & 6.16 & 14.08 & 0.79 & $-$6.73 & 6.22 & 14.36 \\
        7.5 & 19932 & 0.82 & 0.86 & $-$9.67 & 5.65 & 11.41 & 1.00 & $-$9.11 & 5.75 & 11.97 \\
        7.5 & 22021 & 0.95 & 0.76 & $-$11.16 & 5.40 & 9.92 & 0.89 & $-$10.92 & 5.44 & 10.16 \\
\hline
	\end{tabular}
\end{table*}

\begin{table*}
	\centering
	\caption{Same as Tab.~\ref{tab:3d_db_open_models10} but for MLT calibration of open bottom 3D DBA models with $\log{\rm{H}/\rm{He}} = -3$.}
	\label{tab:3d_db_open_models3}
	\begin{tabular}{lcccccccccr} 
		\hline
		$\log{g}$ & $T_{\rm{eff}}$ & 3D~$s_{\rm{env}}$ & \mlts & \q$_{\rm{S}}$ & $(\log{T_{\rm{b}}})_{\rm{S}}$ & $(\log{P_{\rm{b}}})_{\rm{S}}$ &  \mltf & \q$_{\rm{f}}$ & $(\log{T_{\rm{b}}})_{\rm{f}}$ & $(\log{P_{\rm{b}}})_{\rm{f}}$ \\
		& (K) & ($10^9$ erg g$^{-1}$ K$^{-1}$) & & & (K) & (dyn cm$^{-2}$) & & & (K) & (dyn cm$^{-2}$) \\	
		\hline
        7.5 & 11980 & 0.40 & 1.00 & $-$4.56 & 6.59 & 16.54 & 1.00 & $-$4.56 & 6.59 & 16.54 \\
        7.5 & 13855 & 0.44 & 1.00 & $-$4.92 & 6.55 & 16.18 & 1.00 & $-$4.92 & 6.55 & 16.18 \\
        7.5 & 15805 & 0.49 & 1.00 & $-$5.51 & 6.45 & 15.58 & 1.00 & $-$5.51 & 6.45 & 15.58 \\
        7.5 & 18026 & 0.59 & 1.18 & $-$6.66 & 6.23 & 14.42 & 1.38 & $-$6.50 & 6.26 & 14.59 \\
        7.5 & 20035 & 0.80 & 1.08 & $-$9.39 & 5.70 & 11.69 & 1.27 & $-$8.93 & 5.79 & 12.15 \\
        7.5 & 22043 & 0.94 & 0.81 & $-$11.09 & 5.41 & 9.99 & 0.95 & $-$10.85 & 5.45 & 10.23 \\
\hline
	\end{tabular}
\end{table*}

\begin{table*}
	\centering
	\caption{Same as Tab.~\ref{tab:3d_db_open_models10} but for MLT calibration of open bottom 3D DBA models with $\log{\rm{H}/\rm{He}} = -2$.}
	\label{tab:3d_db_open_models2}
	\begin{tabular}{lcccccccccr} 
		\hline
		$\log{g}$ & $T_{\rm{eff}}$ & 3D~$s_{\rm{env}}$ & \mlts & \q$_{\rm{S}}$ & $(\log{T_{\rm{b}}})_{\rm{S}}$ & $(\log{P_{\rm{b}}})_{\rm{S}}$ &  \mltf & \q$_{\rm{f}}$ & $(\log{T_{\rm{b}}})_{\rm{f}}$ & $(\log{P_{\rm{b}}})_{\rm{f}}$ \\
		& (K) & ($10^9$ erg g$^{-1}$ K$^{-1}$) & & & (K) & (dyn cm$^{-2}$) & & & (K) & (dyn cm$^{-2}$) \\	
		\hline
        7.5 & 11977 & 0.45 & 1.00 & $-$5.24 & 6.44 & 15.85 & 1.00 & $-$5.24 & 6.44 & 15.85 \\
        7.5 & 13995 & 0.49 & 1.00 & $-$5.84 & 6.35 & 15.24 & 1.00 & $-$5.84 & 6.35 & 15.24 \\
        7.5 & 16063 & 0.56 & 1.00 & $-$6.97 & 6.15 & 14.11 & 1.00 & $-$6.97 & 6.15 & 14.11 \\
        7.5 & 17963 & 0.68 & 1.00 & $-$8.78 & 5.80 & 12.30 & 1.00 & $-$8.78 & 5.80 & 12.30 \\
        7.5 & 20042 & 0.88 & 1.11 & $-$10.38 & 5.52 & 10.70 & 1.30 & $-$10.23 & 5.55 & 10.85 \\
        7.5 & 21944 & 0.98 & 0.68 & $-$11.49 & 5.34 & 9.59 & 0.80 & $-$11.19 & 5.39 & 9.89 \\
\hline
	\end{tabular}
\end{table*}

\newpage

\newpage

\begin{table*}
	\centering
	\caption{MLT calibration for closed bottom 3D DB models, where $\langle$3D$\rangle$ $T_{\rm{b, \ S}}$ is the \threeD~temperature at the bottom of the \s~boundary, $\langle$3D$\rangle$ $P_{\rm{b, \ S}}$ is the \threeD~pressure at the bottom of the \s~boundary, \mlts~is the calibrated \mlt~value for the \s~boundary and \q$_{\rm{S}}$ is the \q~for the \s~boundary. The same parameters are also given for the flux boundary and are denoted with a subscript `f'.}
	\label{tab:3d_db_modelsclosed10}
	\begin{tabular}{lccccccccccr} 
		\hline
		$\log{g}$ & $T_{\rm{eff}}$ & $\langle$3D$\rangle$ $T_{\rm{b, \ S}}$ & $\langle$3D$\rangle$ $P_{\rm{b, \ S}}$ & \mlts & \q$_{\rm{S}}$ &  $\langle$3D$\rangle$ $T_{\rm{b, \ f}}$ & $\langle$3D$\rangle$ $P_{\rm{b, \ f}}$  & \mltf & \q$_{\rm{f}}$ \\	
		\hline
        7.5 & 26497 & 4.98 & 7.85 & 0.76 & $-$13.20 & 5.10 & 8.28 & 0.85 & $-$12.76 \\
        7.5 & 27993 & 4.90 & 7.45 & 0.69 & $-$13.63 & 4.95 & 7.66 & 0.85 & $-$13.41 \\
        7.5 & 29991 & 4.87 & 7.24 & 0.42 & $-$13.84 & 4.86 & 7.19 & 0.65 & $-$13.82 \\
        7.5 & 32001 & 4.87 & 7.15 & 0.65 & $-$13.91 & 4.85 & 7.08 & 0.65 & $-$13.91 \\        
        \hline
        8.0 & 28107 & 4.99 & 8.21 & 0.65 & $-$13.94 & 5.14 & 8.74 & 0.75 & $-$13.36 \\
        8.0 & 29997 & 4.94 & 7.86 & 0.72 & $-$14.24 & 5.03 & 8.24 & 0.85 & $-$13.84 \\
        8.0 & 31999 & 4.91 & 7.62 & 0.73 & $-$14.47 & 4.94 & 7.76 & 0.89 & $-$14.31 \\
        8.0 & 33999 & 4.89 & 7.43 & 0.65 & $-$14.63 & 4.87 & 7.35 & 0.65 & $-$14.63 \\  
        \hline
        8.5 & 30567 & 5.03 & 8.58 & 0.63 & $-$14.59 & 5.20 & 9.14 & 0.74 & $-$13.96 \\
        8.5 & 32208 & 5.00 & 8.32 & 0.71 & $-$14.80 & 5.12 & 8.77 & 0.81 & $-$14.33 \\
        8.5 & 34020 & 4.95 & 8.00 & 0.75 & $-$15.09 & 5.02 & 8.27 & 0.87 & $-$14.80 \\        
        \hline
        9.0 & 34105 & 5.05 & 8.78 & 0.64 & $-$15.38 & 5.21 & 9.28 & 0.75 & $-$14.81 \\
        \hline
	\end{tabular}
\end{table*}

\begin{table*}
	\centering
	\caption{Same as Tab.~\ref{tab:3d_db_modelsclosed10} but for MLT calibration of closed bottom 3D DBA models with $\log{\rm{H}/\rm{He}} = -7$.}
	\label{tab:3d_modelsclosed7}
	\begin{tabular}{lccccccccccr} 
		\hline
		$\log{g}$ & $T_{\rm{eff}}$ & $\langle$3D$\rangle$ $T_{\rm{b, \ S}}$ & $\langle$3D$\rangle$ $P_{\rm{b, \ S}}$ & \mlts & \q$_{\rm{S}}$ &  $\langle$3D$\rangle$ $T_{\rm{b, \ f}}$ & $\langle$3D$\rangle$ $P_{\rm{b, \ f}}$  & \mltf & \q$_{\rm{f}}$ \\	
		\hline
        7.5 & 26501 & 4.98 & 7.84 & 0.75 & $-$13.21 & 5.11 & 8.29 & 0.85 & $-$12.75 \\
        7.5 & 27993 & 4.90 & 7.46 & 0.70 & $-$13.63 & 4.95 & 7.66 & 0.85 & $-$13.40 \\
        7.5 & 29993 & 4.87 & 7.25 & 0.57 & $-$13.83 & 4.87 & 7.21 & 0.65 & $-$13.82 \\
        7.5 & 32002 & 4.87 & 7.16 & 0.65 & $-$13.91 & 4.85 & 7.06 & 0.65 & $-$13.91 \\     
        \hline
        8.0 & 28037 & 4.98 & 8.18 & 0.64 & $-$13.97 & 5.13 & 8.72 & 0.75 & $-$13.38 \\
        8.0 & 29963 & 5.00 & 8.04 & 0.80 & $-$14.01 & 5.12 & 8.49 & 0.90 & $-$13.55 \\
        8.0 & 32000 & 4.91 & 7.60 & 0.71 & $-$14.48 & 4.94 & 7.73 & 0.86 & $-$14.35 \\
        8.0 & 33999 & 4.89 & 7.43 & 0.65 & $-$14.63 & 4.87 & 7.35 & 0.65 & $-$14.63 \\
        \hline
        8.5 & 30514 & 5.03 & 8.59 & 0.63 & $-$14.58 & 5.21 & 9.15 & 0.74 & $-$13.94 \\
        8.5 & 32012 & 4.99 & 8.33 & 0.70 & $-$14.80 & 5.14 & 8.83 & 0.80 & $-$14.26 \\
        8.5 & 33949 & 4.95 & 8.00 & 0.74 & $-$15.09 & 5.02 & 8.28 & 0.87 & $-$14.79 \\ 
        \hline
        9.0 & 34021 & 5.06 & 8.82 & 0.65 & $-$15.33 & 5.21 & 9.29 & 0.75 & $-$14.79 \\
        \hline
	\end{tabular}
\end{table*}

\begin{table*}
	\centering
	\caption{Same as Tab.~\ref{tab:3d_db_modelsclosed10} but for MLT calibration of closed bottom 3D DBA models with $\log{\rm{H}/\rm{He}} = -5$.}
	\label{tab:3d_modelsclosed5}
	\begin{tabular}{lccccccccccr} 
		\hline
		$\log{g}$ & $T_{\rm{eff}}$ & $\langle$3D$\rangle$ $T_{\rm{b, \ S}}$ & $\langle$3D$\rangle$ $P_{\rm{b, \ S}}$ & \mlts & \q$_{\rm{S}}$ &  $\langle$3D$\rangle$ $T_{\rm{b, \ f}}$ & $\langle$3D$\rangle$ $P_{\rm{b, \ f}}$  & \mltf & \q$_{\rm{f}}$ \\	
		\hline
        7.5 & 26522 & 4.97 & 7.83 & 0.75 & $-$13.23 & 5.10 & 8.27 & 0.85 & $-$12.77 \\
        7.5 & 27998 & 4.90 & 7.46 & 0.70 & $-$13.63 & 4.95 & 7.66 & 0.85 & $-$13.41 \\
        7.5 & 29992 & 4.87 & 7.25 & 0.49 & $-$13.84 & 4.86 & 7.19 & 0.65 & $-$13.82 \\
        7.5 & 32002 & 4.87 & 7.16 & 0.65 & $-$13.91 & 4.85 & 7.08 & 0.65 & $-$13.91 \\
        \hline
        8.0 & 28086 & 4.98 & 8.19 & 0.65 & $-$13.95 & 5.13 & 8.73 & 0.75 & $-$13.38 \\
        8.0 & 29989 & 4.94 & 7.87 & 0.72 & $-$14.24 & 5.03 & 8.24 & 0.85 & $-$13.84 \\
        8.0 & 32002 & 4.91 & 7.61 & 0.71 & $-$14.48 & 4.94 & 7.75 & 0.88 & $-$14.32 \\
        8.0 & 34000 & 4.89 & 7.43 & 0.65 & $-$14.63 & 4.87 & 7.35 & 0.65 & $-$14.63 \\      
        \hline
        8.5 & 30517 & 5.02 & 8.59 & 0.63 & $-$14.58 & 5.21 & 9.16 & 0.74 & $-$13.93 \\
        8.5 & 32015 & 5.00 & 8.36 & 0.71 & $-$14.76 & 5.14 & 8.83 & 0.80 & $-$14.27 \\
        8.5 & 33947 & 4.95 & 8.00 & 0.74 & $-$15.09 & 5.02 & 8.29 & 0.87 & $-$14.78 \\      
        \hline
        9.0 & 33986 & 5.06 & 8.82 & 0.65 & $-$15.33 & 5.21 & 9.30 & 0.75 & $-$14.79 \\
        \hline
	\end{tabular}
\end{table*}

\begin{table*}
	\centering
	\caption{Same as Tab.~\ref{tab:3d_db_modelsclosed10} but for MLT calibration of closed bottom 3D DBA models with $\log{\rm{H}/\rm{He}} = -4$.}
	\label{tab:3d_modelsclosed4}
	\begin{tabular}{lccccccccccr} 
		\hline
		$\log{g}$ & $T_{\rm{eff}}$ & $\langle$3D$\rangle$ $T_{\rm{b, \ S}}$ & $\langle$3D$\rangle$ $P_{\rm{b, \ S}}$ & \mlts & \q$_{\rm{S}}$ &  $\langle$3D$\rangle$ $T_{\rm{b, \ f}}$ & $\langle$3D$\rangle$ $P_{\rm{b, \ f}}$  & \mltf & \q$_{\rm{f}}$ \\	
			& (K) & (K) & (dyn cm$^{-2}$) & & & (K) & (dyn cm$^{-2}$) & \\	
		\hline
        7.5 & 26632 & 4.97 & 7.81 & 0.76 & $-$13.24 & 5.09 & 8.23 & 0.86 & $-$12.81 \\
        7.5 & 28004 & 4.90 & 7.45 & 0.69 & $-$13.64 & 4.94 & 7.64 & 0.84 & $-$13.43 \\
        7.5 & 29993 & 4.87 & 7.25 & 0.51 & $-$13.84 & 4.86 & 7.19 & 0.65 & $-$13.82 \\
        7.5 & 32002 & 4.87 & 7.15 & 0.65 & $-$13.91 & 4.85 & 7.07 & 0.65 & $-$13.91 \\     
		\hline 
        8.0 & 28092 & 4.98 & 8.17 & 0.64 & $-$13.97 & 5.12 & 8.68 & 0.74 & $-$13.42 \\
        8.0 & 29994 & 4.94 & 7.87 & 0.72 & $-$14.23 & 5.04 & 8.25 & 0.85 & $-$13.82 \\
        8.0 & 32003 & 4.91 & 7.60 & 0.71 & $-$14.48 & 4.94 & 7.73 & 0.86 & $-$14.35 \\
        8.0 & 34000 & 4.89 & 7.43 & 0.65 & $-$14.63 & 4.87 & 7.35 & 0.65 & $-$14.63 \\
        \hline
        8.5 & 30490 & 5.03 & 8.61 & 0.63 & $-$14.56 & 5.20 & 9.15 & 0.73 & $-$13.95 \\
        8.5 & 32008 & 5.00 & 8.35 & 0.71 & $-$14.77 & 5.13 & 8.81 & 0.80 & $-$14.29 \\
        8.5 & 33963 & 4.95 & 8.00 & 0.75 & $-$15.09 & 5.01 & 8.27 & 0.87 & $-$14.80 \\
        \hline
        9.0 & 33988 & 5.06 & 8.83 & 0.65 & $-$15.33 & 5.21 & 9.30 & 0.75 & $-$14.79 \\
        \hline
	\end{tabular}
\end{table*}

\begin{table*}
	\centering
	\caption{Same as Tab.~\ref{tab:3d_db_modelsclosed10} but for MLT calibration of closed bottom 3D DBA models with $\log{\rm{H}/\rm{He}} = -3$.}
	\label{tab:3d_modelsclosed3}
	\begin{tabular}{lccccccccccr} 
		\hline
		$\log{g}$ & $T_{\rm{eff}}$ & $\langle$3D$\rangle$ $T_{\rm{b, \ S}}$ & $\langle$3D$\rangle$ $P_{\rm{b, \ S}}$ & \mlts & \q$_{\rm{S}}$ &  $\langle$3D$\rangle$ $T_{\rm{b, \ f}}$ & $\langle$3D$\rangle$ $P_{\rm{b, \ f}}$  & \mltf & \q$_{\rm{f}}$ \\	
			& (K) & (K) & (dyn cm$^{-2}$) & & & (K) & (dyn cm$^{-2}$) & \\
		\hline
        7.5 & 26670 & 4.95 & 7.74 & 0.74 & $-$13.34 & 5.06 & 8.15 & 0.85 & $-$12.91 \\
        7.5 & 28000 & 4.90 & 7.44 & 0.69 & $-$13.65 & 4.94 & 7.61 & 0.83 & $-$13.45 \\
        7.5 & 29999 & 4.87 & 7.24 & 0.54 & $-$13.84 & 4.85 & 7.15 & 0.65 & $-$13.82 \\
        7.5 & 32000 & 4.87 & 7.15 & 0.65 & $-$13.92 & 4.84 & 7.05 & 0.65 & $-$13.92 \\   
        \hline
        8.0 & 28118 & 4.98 & 8.14 & 0.64 & $-$13.99 & 5.11 & 8.64 & 0.74 & $-$13.45 \\
        8.0 & 30001 & 4.94 & 7.83 & 0.71 & $-$14.27 & 5.02 & 8.18 & 0.84 & $-$13.90 \\
        8.0 & 31999 & 4.90 & 7.58 & 0.67 & $-$14.51 & 4.93 & 7.68 & 0.82 & $-$14.40 \\
        8.0 & 33980 & 4.89 & 7.43 & 0.65 & $-$14.63 & 4.87 & 7.36 & 0.65 & $-$14.63 \\
        \hline
        8.5 & 30364 & 5.03 & 8.59 & 0.62 & $-$14.58 & 5.20 & 9.14 & 0.72 & $-$13.96 \\
        8.5 & 31965 & 4.99 & 8.30 & 0.69 & $-$14.81 & 5.12 & 8.77 & 0.79 & $-$14.31 \\
        8.5 & 34038 & 4.95 & 7.99 & 0.75 & $-$15.10 & 5.01 & 8.26 & 0.88 & $-$14.81 \\
        \hline
        9.0 & 33917 & 5.06 & 8.84 & 0.65 & $-$15.31 & 5.21 & 9.31 & 0.75 & $-$14.77 \\
        \hline
	\end{tabular}
\end{table*}

\begin{table*}
	\centering
	\caption{Same as Tab.~\ref{tab:3d_db_modelsclosed10} but for MLT calibration of closed bottom 3D DBA models with $\log{\rm{H}/\rm{He}} = -2$.}
	\label{tab:3d_modelsclosed2}
	\begin{tabular}{lccccccccccr} 
		\hline
		$\log{g}$ & $T_{\rm{eff}}$ & $\langle$3D$\rangle$ $T_{\rm{b, \ S}}$ & $\langle$3D$\rangle$ $P_{\rm{b, \ S}}$ & \mlts & \q$_{\rm{S}}$ &  $\langle$3D$\rangle$ $T_{\rm{b, \ f}}$ & $\langle$3D$\rangle$ $P_{\rm{b, \ f}}$  & \mltf & \q$_{\rm{f}}$ \\	
			& (K) & (K) & (dyn cm$^{-2}$) & & & (K) & (dyn cm$^{-2}$) & \\	
		\hline
        7.5 & 26471 & 4.93 & 7.67 & 0.71 & $-$13.43 & 5.02 & 8.04 & 0.84 & $-$13.04 \\
        7.5 & 27996 & 4.89 & 7.38 & 0.67 & $-$13.70 & 4.91 & 7.50 & 0.80 & $-$13.57 \\
        7.5 & 29982 & 4.87 & 7.22 & 0.65 & $-$13.84 & 4.85 & 7.13 & 0.65 & $-$13.84 \\
        7.5 & 32009 & 4.86 & 7.12 & 0.65 & $-$13.93 & 4.84 & 7.04 & 0.65 & $-$13.93 \\
        \hline
        8.0 & 27968 & 4.96 & 8.07 & 0.65 & $-$14.05 & 5.09 & 8.54 & 0.75 & $-$13.56 \\
        8.0 & 30013 & 4.93 & 7.77 & 0.71 & $-$14.32 & 4.99 & 8.06 & 0.86 & $-$14.01 \\
        8.0 & 31998 & 4.90 & 7.53 & 0.61 & $-$14.55 & 4.91 & 7.58 & 0.74 & $-$14.50 \\
        8.0 & 33989 & 4.88 & 7.41 & 0.65 & $-$14.65 & 4.86 & 7.33 & 0.65 & $-$14.65 \\
        \hline
        8.5 & 30535 & 5.01 & 8.45 & 0.64 & $-$14.69 & 5.17 & 8.99 & 0.75 & $-$14.10 \\
        8.5 & 31852 & 4.98 & 8.25 & 0.70 & $-$14.86 & 5.10 & 8.70 & 0.81 & $-$14.39 \\
        8.5 & 33930 & 4.94 & 7.95 & 0.75 & $-$15.14 & 5.00 & 8.20 & 0.89 & $-$14.87 \\
        \hline
        9.0 & 33770 & 5.05 & 8.78 & 0.66 & $-$15.36 & 5.20 & 9.27 & 0.76 & $-$14.81 \\
        \hline
	\end{tabular}
\end{table*}



\label{lastpage}
\end{document}